\documentclass{emulateapj}
\slugcomment{{\sc Accepted to ApJ:} June 24, 2011}
\usepackage{wasysym} 
\usepackage{graphicx}
\usepackage{subfigure}
\newcommand{\hii}{H\,{\sc ii}}
\newcommand{\mum}{\ensuremath{\mu\mbox{m}}}

\begin{document}

\title{The Physical Conditions in Starbursts derived from Bayesian fitting of mid-IR SED\lowercase{s}: 30 Doradus as a Template} 

\shorttitle{30 Dor SED}

\author{J.~R. Mart\'{\i}nez-Galarza\altaffilmark{1}, B. Groves\altaffilmark{1,2}, B. Brandl\altaffilmark{1}, G. E. de Messieres\altaffilmark{3}, R. Indebetouw\altaffilmark{3}, M.~A. Dopita\altaffilmark{4,5,6}
}

\altaffiltext{1}{Leiden Observatory, Leiden University, P.O. Box 9513, 2300 CA Leiden, The Netherlands}
\altaffiltext{2}{Max Planck Institute for Astronomy, K\"{o}nigstuhl 17, D-69117, Heidelberg, Germany}
\altaffiltext{3}{University of Virginia, Charlottesville, VA USA}
\altaffiltext{4}{Mount Stromlo and Siding Spring Observatories, Research School of Astronomy and Astrophysics, Australian National University, Cotter Road, Weston Creek, ACT 2611, Australia.}
\altaffiltext{5}{Astronomy Department, King Abdulaziz University, P.O. Box 80203, Jeddah, Saudi Arabia}
\altaffiltext{6}{Institute for Astronomy, University of Hawaii, 2680  Woodlawn Drive, Honolulu, HI 96822}

\begin{abstract}

To understand and interpret the observed Spectral Energy Distributions (SEDs) of starbursts, theoretical or semi-empirical SED models are necessary. Yet, while they are well-founded in theory, independent verification and calibration of these models, including the exploration of possible degeneracies between their parameters, are rarely made. As a consequence, a robust fitting method that leads to unique and reproducible results has been lacking.  Here we introduce a novel approach based on Bayesian analysis to fit the Spitzer-IRS spectra of starbursts using the SED models proposed by \citet{Groves08}.  We demonstrate its capabilities and verify the agreement between the derived best fit parameters and actual physical conditions by modelling the nearby, well-studied, giant \hii\ region 30~Dor in the LMC.  The derived physical parameters, such as cluster mass, cluster age, ISM pressure and covering fraction of photodissociation regions, are representative of the 30~Dor region.  The inclusion of the emission lines in the modelling is crucial to break degeneracies.  We investigate the limitations and uncertainties by modelling sub-regions, which are dominated by single components, within 30~Dor.  A remarkable result for 30 Doradus in particular is a considerable contribution to its mid-infrared spectrum from hot ($\approx 300\: $K) dust. The demonstrated success of our approach will allow us to derive the physical conditions in more distant, spatially unresolved starbursts.

\end{abstract}

\keywords{SED fitting -- \hii\ regions -- infrared: ISM -- ISM: individual: 30 Dor -- stars: formation}

\section{Introduction}
\label{sec:introduction}

In theory, the spectral energy distribution (SED) of a galaxy contains a wealth of information about both its evolutionary history and current conditions. 
However, extracting this information is difficult and requires the use of physically based models. Nevertheless, SED fitting is a necessary process as many high redshift galaxies remain unresolved by our current instruments and any attempts to characterize the conditions and processes that lead to their starburst activities rely almost exclusively on their spatially averaged properties. These models of the integrated SEDs of galaxies currently cover a wide range of galaxy types, but are particularly dominated by models of Starburst galaxies \citep{Galliano03, Siebenmorgen07, Takagi03, Silva98, Dopita05, Dopita06b, Dopita06c, Groves08}. The ultraviolet (UV) to far infrared (FIR) SED of these \textit{Starbursts} is dominated by the energetic photons emitted by massive stars with typical lifetimes of less than 10$\: $Myr.

In particular, the mid-infrared portion of the SED contains several important diagnostics that probe the physical conditions of starbursts.Observations of a set of marginally resolved starburst galaxies with the Spitzer Space Telescope show a broad range of mid-infrared properties, including different strengths of the polycyclic aromatic hydrocarbon (PAH) bands, thermal continuum slopes, depth of the silicate absorption features at 10$\: $\mum\ and 18$\: $\mum\ and intensity of nebular emission lines \citep{Brandl06, Bernard_Salas09}. All these signatures have contributions from different spatial regions, depending on the geometrical distribution of gas and dust with respect to the ionizing stars. For example, \cite{Beirao09} reported on the presence of compact star forming knots around the nucleus of the starburst galaxy Arp 143, and similar star forming knots have been reported near the nucleus of NGC 253 \citep{Fernandez_Ontiveros09}. In other galaxies, such as M51, star formation spreads more uniformly over the galactic disk. The different distributions of gas, dust, and stars in galaxies affect the shape of the spatially integrated SED. Inversely, a sophisticated and well calibrated SED model should be able to recover the information on the local starburst conditions from the integrated SED.

A considerable amount of SED model libraries can be found in recent literature \citep[see e.g.][for a comprehensive review on SED fitting]{Walcher11}.  These models generally make assumptions on the internal physics of galaxies and predict the output SED as a function of certain model parameters, such as star formation rates (SFRs), metallicity (Z), and the interstellar medium (ISM) pressure, density, and temperature, among many others. SED fitting refers to the process of choosing from a particular library the model solutions that best reproduce the data. While finding the best-fit model via, for an example, a $\chi^2$ minimization provides an estimate of the parameters, this method alone is insufficient to provide absolute parameter uncertainties. In order to obtain robust parameter estimates, including uncertainties, it becomes necessary to explore the whole parameter space and perform a statistical study of their correlations. We highlight four aspects that make this task difficult. First, the sensitivity of photometric and spectroscopic studies is limited not only by instrumental constraints, but also by more fundamental constrains such as shot noise in the case of weak sources. Hence, the robustness of SED fitting depends on the data quality and on sufficient data coverage. Second, degeneracies between model parameters are common, especially when limited to a narrow spectral window (e.g., the mid-infrared). Third, independent determinations (from observations or theory) of the physical parameters against which we can confront our model results are rare for most starburst, hence making it difficult to calibrate the models. And last but not least, no robust fitting routine that leads to reproducible results has been established so far for the specific case of starburst spectra.

In this paper we present a Bayesian fitting routine for the mid-infrared ($5-38\: $\mum) spectra of starbursts that can be extended to other wavelengths. We derive probability distribution functions (PDFs) for the model parameters, and study the implications on the physics of starbursts. To calibrate this routine we apply it to the mid-infrared spectrum of the 30 Doradus region in the Large Magellanic Cloud. The selection of this nearby starburst as a calibrator is natural, since its proximity ($\approx 53\: \rm{pc}$) allows us to differentiate spatially resolved sub-regions of the giant \hii\ region, and study their spectra separately. The well studied stellar populations, ionized gas, and dust content provide the necessary independent measurements to compare with SED fitting results. 

Current spatial resolutions achieved with the mapping mode of the \textit{Infrared Spectrograph} on board the Spitzer Space Telescope are of the order of a few arcseconds at 5$\: $\mum\, corresponding to a scale of about one parsec at the distance to 30 Doradus. Even the next generation spectrometer operating at these wavelengths, the \textit{Mid Infrared Instrument} (MIRI), on board the 6.5$\: $m James Webb Space Telescope, will not be able to resolve typical giant \hii\ regions in galaxies located at distances larger than about 30~Mpc at a nominal wavelength of 15$\: $\mum. This highlights the importance of understanding the integrated SEDs of these objects.

This paper is structured as follows. In \S \ref{sec:data} we describe some general aspects of the 30 Doradus region, focusing on its stellar content and its physical properties, as obtained from HST and Spitzer observations, and we discuss the Spitzer-IRS spectral data that we model. In \S \ref{sec:models} we give a brief overview of the models we use to generate our grid of synthetic SEDs. In \S \ref{sec:fitting} we introduce our fitting routine and discuss the assumed priors and involved uncertainties. \S \ref{sec:results} presents the results of applying our fitting routine to 30 Doradus, discuss the implications of the model parameters and the physical interpretation of the mid-infrared SEDs. Finally, in \S \ref{sec:discusion} we summarize our main findings.

\section{The 30 Doradus Region}
\label{sec:data}

Our choice of 30 Doradus as a calibrator relies on three powerful reasons: \textit{(i)} it is the largest giant \hii\ region in the Local Group, \textit{(ii)} it is well studied across the whole electromagnetic spectrum, and \textit{(iii)} it is close enough to be well resolved into individual components. In this section we describe the general properties of 30 Doradus and the spectral data that we model.

\subsection{Properties of the 30 Doradus Region}
\label{sec:properties}

\begin{figure*}[ht] \epsscale{1.0} \begin{center} \rotatebox{0}{\plotone{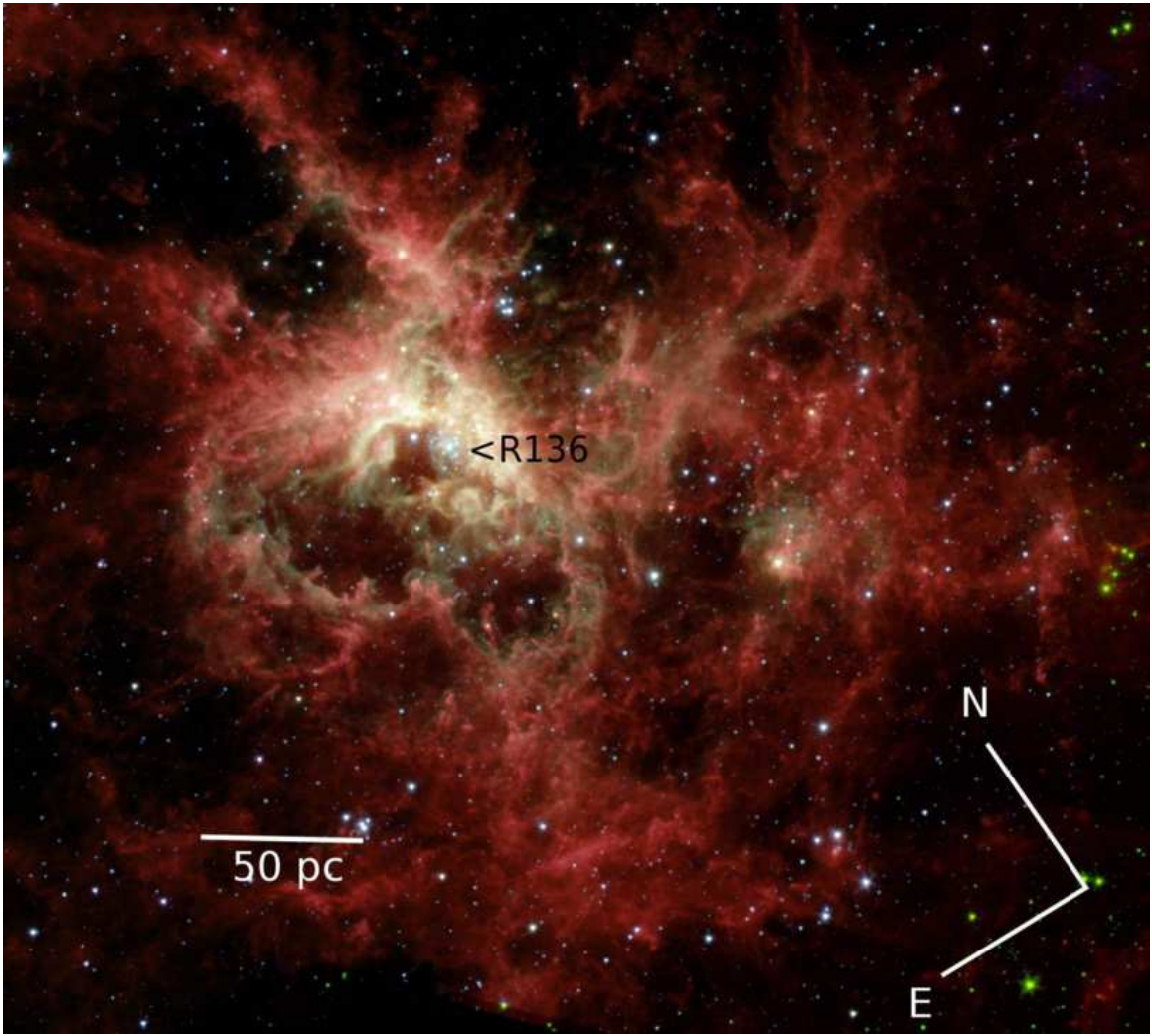}} \end{center} \caption{\label{fig:30dor_overview} The 30 Doradus region imaged in the 4 Spitzer-IRAC channels. The filamentary structure and bubble-like cavities are evident. The ionized gas illuminated by R136 (green) is confined to a thin layer next to the PDR (red), where the PAH emission is found.} \end{figure*}

30 Doradus is the most massive giant \hii\ region in the Local Group. It is located $53 \pm 3\: $kpc away \citep{Feast97}, in the north-east part of the Large Magellanic Cloud (LMC) and includes the stellar cluster NGC 2070, the cloud of ionized gas created by the ionizing radiation from NGC 2070 and dominated by its compact central core R136,and the photon-dissociated regions and molecular material associated with the star forming region. We show the complexity of the region in Fig.~\ref{fig:30dor_overview}. 

R136 is the most dense concentration of stars in the local group, with an estimated stellar mass of $2\times 10^4\: \rm{M}_{\astrosun}$ contained within the innermost 5$\: $pc \citep{Hunter95}. The associated \hii\ region has an H$\alpha$ luminosity of $1.5\times10^{40}\: $erg$\: $s$^{-1}$ \citep{Kennicutt84} and a far-infrared luminosity of $4\times10^7\: \rm{L}_{\astrosun}$ \citep{Werner78}. Stellar winds, supernovae, and radiation pressure from the central cluster have excavated an expanding ionized bubble and created a complex filamentary structure (Fig.~\ref{fig:30dor_overview}). This bubble, and other similar cavities in the region are filled with X-ray emitting gas at temperatures of $\sim10^6\: $K, as revealed by observations with the Chandra Space Observatory \citep{Townsley06}. A recent study of the optical emission lines shows no evidence of ionization by supernova-driven shocks found by a recent study\citep{Pellegrini10}, and hence the dominant excitation mechanism in the 30 Doradus region is photoionization by the UV photons produced mainly in R136. This was corroborated by a comparison of observed IRS line fluxes with models of the mid-infrared lines \citep{Indebetouw09}.

Using HST spectroscopy, \cite{Walborn97} identified several non-coeval stellar populations in the 30 Doradus region, and classified them as follows: \textit{(i)} a core-ionizing phase (R136), with an age of 2-3~Myr; \textit{(ii)} a peripheral triggered phase, with an age of $<1$~Myr \citep[this population has also been identified using near infrared excess measurements, e.g.][]{Maercker05}; \textit{(iii)} a phase of OB supergiants with an age of 6~Myr; \textit{(iv)} the Hodge 301 cluster, $3^{\prime}$ NW of R136, with an age of $\approx 10$~Myr, and \textit{(v)} the R143 OB association, with ages between 4-7~Myr.

An interesting aspect of 30 Doradus is its structure of bubbles and filaments. Observations of galactic and extragalactic \ion{H}{2} regions have revealed expanding structures of ionized gas driven by stellar winds and supernova activity from the OB stellar population. In the particular case of 30 Doradus, expanding supershells have been detected with diameters between 2 and 20$\: $pc and expansion velocities of 100-300$\: $km$\: $s$^{-1}$ \citep{Chu94}. 

The metallicity of 30 Doradus and of the LMC in general is sub-solar ($Z = 0.4\: Z_{\astrosun}$) \citep{Westerlund97}. Due to this low metallicity environment, the dust-to-gas ratio in the LMC is about 30\% lower than in the Milky Way \citep[see review by][and references therein]{Draine03}, and the system allows us to investigate the effect of UV radiation in lower metallicity environments as compared to our own galaxy.

For simplicity, in this paper we refer to 30 Doradus as the region of $\approx 100\: $pc$=4.1\: $arcmin in diameter in projection centered in R136.

\subsection{The Integrated Mid-IR Spectrum of 30 Doradus}
\label{sec:spectral_map}

The Spitzer-IRS spectral data that we model here has been extensively discussed in \cite{Indebetouw09}, as part of the Spitzer General Observer Program \textit{Stellar Feedback on Circumcluster Gas and Dust in 30 Doradus, the Nearest Super-Star Cluster}, (PID 30653, P.~I. R. Indebetouw). It consists of four data cubes obtained by mapping the 30 Doradus region with the two low-resolution slits of the IRS (``short-low'' and ``long-low'') in each of their two spectral orders. For reference, the first order of the short-low (SL1) map covers an area of $116\: \rm{pc} \times 84\: $pc, and includes a significant portion of the 30 Doradus emission nebula. The wavelength coverage is between 5-38$\: $\mum\ with a resolving power $R=\lambda/\Delta\lambda$, varying from 60 at the short wavelength end to about 110 at the long wavelength end. Exposure times were of the order of 150$\: $s per slit position.

\begin{figure}[ht] \epsscale{0.85} \begin{center} \rotatebox{270}{\plotone{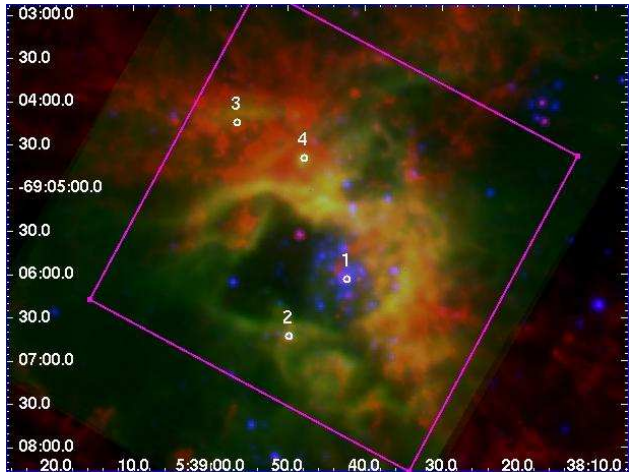}} \end{center} \caption{\label{fig:30dor_rgb_map} Multi-wavelength view of the 30 Doradus region. Red: IRAC 8$\: $\mum\ image showing the PAH emission from the PDR region (c.f.~\ref{fig:30dor_overview}). Green: [\ion{S}{4}]10.5$\: $\mum\ emission line map, constructed from the spectral map described in \S \ref{sec:spectral_map}, tracing the distribution of highly ionized gas. Blue: Red continuum image showing the stellar continuum emission. White circles mark the positions of the individual spectra discussed in \S \ref{sec:locations}, and their sizes correspond to the size of one resolution element of the spectral map. The magenta square outlines the full IRS spectral map explored in this paper. North is up and east is to the left.} \end{figure}

\begin{figure}[ht] \epsscale{1.2} \begin{center} \rotatebox{0}{\plotone{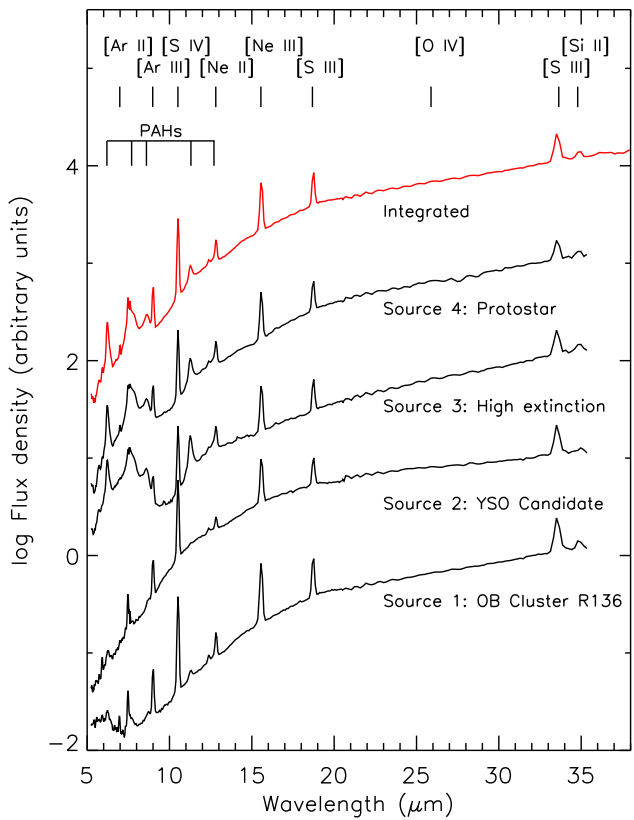}} \end{center} \caption{\label{fig:all_spectra} Integrated mid-infrared spectrum of 30 Doradus (red line) and spectra of sources described in \S \ref{sec:locations} (black lines) as labelled. All spectra are normalized to the flux at 30$\: $\mum\ and are shifted one decade in flux for comparison. The main spectral features are labelled.} \end{figure}

Spectra of chosen regions are extracted using the CUBISM software package \citep{Smith07}. Once the sky subtraction has been performed, we extract individual spectra using a resolution element of $2\times 2$ SL1 pixels for all orders. This corresponds to an angular resolution of $3.7$ arcseconds, and a physical spatial resolution of roughly 1~pc at the distance to the LMC. To create the spatially integrated spectrum of 30 Doradus, we co-add the spectra of all individual resolution elements within an area of about $64\: \rm{pc} \times 63\: $pc (the magenta square in Fig.~\ref{fig:30dor_rgb_map}). We show the resulting integrated spectrum in Fig.~\ref{fig:all_spectra}. The integrated spectrum is dominated by emission from nebular lines and the thermal continuum, while the PAH emission is generally weak in the region.

Here we express all fluxes as $\nu F_{\nu}$ in units of erg$\: $s$^{-1}$. To convert from the MJy$\: \rm{sr}^{-1}$ units from the IRS pipeline, we multiply the fluxes by the aperture area of 13.7 arcsec$^2$, and assume a distance to 30 Doradus (LMC) of $53\: $kpc \citep{Feast97}.

\subsection{Individual Sources}
\label{sec:locations}

In Fig.~\ref{fig:30dor_rgb_map} we have indicated four locations defined in Table \ref{tab:regions}, of which we show their respective spectra in Fig.~\ref{fig:all_spectra}. These locations include sources of different nature and were chosen to cover a broad range of physical conditions and spectral shapes. We model their spectra separately to study the validity of the models in environments which are dominated by either highly ionized gas or by embedded stars.

Source 1 corresponds to the location of the young OB cluster R136. The emission here is dominated by UV and optical photons and shows little infrared emission from PAHs. 

Source 2 is a YSO candidate selected from IRAC colors \citep{Kim07}, according to the criterion suggested by \cite{Allen04}, about 1 arcminute southwest of R136, at the ionized southern edge of the main bubble-like structure, in a region with significant [\ion{S}{4}]10.5$\: $\mum\ emission. Its spectrum has a smooth thermal continuum with no sign of PAH emission, bu with the typical nebular lines [\ion{Ne}{2}]12.81$\: $\mum, [\ion{Ne}{3}]15.56$\: $\mum, and [\ion{S}{3}]18.71$\: $\mum. 

Source 3 is a bright infrared source  outside of the main bubble, to the north-west of the cluster. Its spectrum shows prominent PAH emission features and a deep silicate absorption feature at 10$\: $\mum. 

Source 4 is an infrared source identified as a protostellar object by \cite{Walborn87}, just outside the main bubble, north of R136. It coincides with a strong peak of [\ion{S}{4}] emission and is also an X-ray source. \cite{Lazendic03} even consider this source to be a supernova remnant, but also point to its higher H$\alpha$/H$\beta$ ratio and the possibility of it being an \hii\ region with an extinction higher than average.

\begin{deluxetable*}{lccc} 
\tablecolumns{4}
\tablecaption{Localized sub-regions in the 30 Doradus Spectral Map.}

\tablehead{
  \colhead{Object} &
  \colhead{RA} &
  \colhead{Dec} &
  \colhead{Remarks} 
}
 \startdata
 Source 1  & $5^{\rm{h}}\: 38^{\rm{m}}\: 42.3^{\rm{s}}$ & $-69^\circ \: 06^{\prime} \: 03.0^{\prime\prime}$ & R136\\
 Source 2  &  $5^{\rm{h}}\: 38^{\rm{m}}\: 49.7^{\rm{s}}$ & $-69^\circ \: 06^{\prime} \: 42.7^{\prime\prime}$ & YSO candidate\\
 Source 3  &  $5^{\rm{h}}\: 38^{\rm{m}}\: 56.5^{\rm{s}}$ & $-69^\circ \: 04^{\prime} \: 16.9^{\prime\prime}$ & High extinction ($\tau_{10\: \mu \rm{m}} \approx 0.60$)\\
 Source 4  & $5^{\rm{h}}\: 38^{\rm{m}}\: 48.30^{\rm{s}}$ & $-69^\circ \: 04^{\prime} \: 41.2^{\prime\prime}$ & Protostar, [\ion{S}{4}] emission\\
 
 \enddata
\label{tab:regions}
\end{deluxetable*}

In general, we observe that  emission from all PAH bands is weak towards 30 Doradus as compared to other starburst systems (see, for example the starburst SED template in \citet{Brandl06}). In particular, the 17$\: $\mum\ PAH complex generally associated with out-of-band bending modes of large neutral PAH grains \citep{Kerckhoven00, Peeters04} is only marginally detected in our spectra. A remarkable result regarding this point is that the 17$\: $\mum\ complex is weaker towards source 3 than expected from the proportionality relations that have been empirically derived between different PAH bands \citep{Smith07}. This proportionality implies that in starburst galaxies the equivalent width of the 11.3$\: $\mum\ feature is about twice the equivalent width of the 17$\: $\mum\ feature \citep{Brandl06}. If this were to hold also for our source 3, we would expect a flux density of the 17$\: $\mum\ 20\% higher than the thermal continuum at this wavelength. However, our data indicates an upper limit for the 17$\: $\mum\ emission of only 2\% above the continuum level.

This suppression of the 17$\: $\mum\ band can have several interpretations. A possibility is that the PAH molecules are not neutral in this region of 30 Doradus. However, source 3 is outside of the main ionized bubble shown in Fig. \ref{fig:30dor_rgb_map}, and hence we do not expect a high ionization state of the PAHs in this region. Metallicity variations could also account for a change in the relative strength of the 17$\: $\mum\ feature \citep{Smith07}, but even in very low metallicity environments an extremely weak 17$\: $\mum\ would also imply a weak 11.3$\: $\mum\ feature, which we do not observe. We are left with the explanation of grain size effects. As mentioned, emission features between 15$\: $\mum\ and 20$\: $\mum\ are associated with large PAH grains, typically containing $\approx 2000$ carbon atoms \citep{Kerckhoven00}. Wether the conditions in 30 Doradus are unfavorable for the formation of large PAH grains is the mater of a subsequent paper.

\section{Modelling the SEDs of Starbursts}
\label{sec:models}

\subsection{Literature on SED modelling}
\label{sec:literature}

The simplest Spectral Energy Distribution models consider a starburst as a single spherical \hii\ region surrounding a central ionizing cluster, use stellar synthesis for the stellar radiation and solve the radiative transfer for dust and gas in spherical geometry. These semi-empirical attempts use observations of specific objects, such as star-forming dwarf galaxies \citep{Galliano03} or nuclear starbursts  \citep{Siebenmorgen07} to constrain the model parameters. They are successful in reproducing the photometry, and to some extent the IR spectra of these objects, but are limited to a narrow range of physical conditions (e.g., only two orders of magnitude in dust density).  Fully theoretical models, such as the ones proposed by \cite{Takagi03}, make similar assumptions on geometry, dust properties and stellar synthesis, and cover a broader range of physical properties to model a larger sample of starburst galaxies, but ignore spatial variations of the parameters. 

More sophisticated models consider the starburst as a collection of individual \hii\ regions with different ages and environments, whose SEDs add up to produce the total galactic SED. In the GRASIL models, for example, each of these individual \ion{H}{2} regions is assumed to have different physical properties \citep{Silva98}. Unfortunately, they do not allow for the  dynamical evolution of the expanding shell-like structures such as the ones we have described in \S \ref{sec:properties}. In the expanding mass-loss bubble scenario, the time-dependent radius and external pressure of the \hii\ region are controlled by the mechanical luminosity from the newborn stars \citep{Castor75}, and have a strong influence on the shape of the SED, as they control the gas and dust geometry \citep{Groves08}.

None of the existing starburst models simultaneously accounts for both the multiplicity of \hii\ regions in a starburst system and their time evolution as individual \hii\ regions evolve as mass-losing bubbles. However, the models described in the series of papers \cite{Dopita05}, \cite{Dopita06b}, \cite{Dopita06c} and \cite{Groves08}  (D\&G models hereafter), represent a step forward in our theoretical description of starburst systems, by including these two aspects in a self-consistent way. Although these models have been successfully applied to the SEDs of a variety of objects, such as brightest cluster galaxies (BCGs) \citep{Donahue11}, no systematic study of the model degeneracies have been presented. In the remainder of this section we briefly describe the underlying physics of the D\&G models, emphasizing the aspects that are relevant for our discussion, and connect this description to the controlling model parameters. For a detailed description of the model, we refer the reader to the Dopita \& Groves paper series.

\subsection{The physical concept behind the model}
\label{sec:models_brent}

The D\&G models compute the SED of a starburst galaxy as the sum of the SEDs of individual expanding \hii\ regions, averaged over ages younger than 10$\: $Myr. By this age, over $95\%$ of the total ionizing photons produced during the main sequence stage of the massive stars have been emitted \citep[e.g.][]{Dopita06b} and the non-ionizing UV flux is rapidly decreasing as the OB stars evolve off the main sequence into supernovae. Here we will test the applicability of the individual \hii\ regions that constitute the building blocks of the models. Each individual giant \hii\ region evolves in time as a bubble expanding into the surrounding ISM, driven by the stellar winds and supernova from the central cluster. The dynamical evolution is controlled by the equations of motion of the expanding bubble \citep{Castor75} and provides the instantaneous distance of the ionization front, dust and molecular gas with respect to the central cluster. The time-dependent expansion of this mass-loss bubble controls the temperature of the dust and the ionization state of the gas in the \hii\ region, altering the shape of the SED. 

The stellar synthesis code Starburst99 \citep{Leitherer99, Vazquez05} provides the stellar radiation field for a population of stars at a given age and metallicity. The energy output is normalized to a template cluster, whose mass is a free parameter of the models and can be scaled to any desired value. The stellar mass in the cluster is distributed according to a Kroupa IMF with a lower cutoff at 0.1$\: \rm{M}_{\astrosun}$ and an upper cutoff of 120$\: \rm{M}_{\astrosun}$ \citep{Kroupa02}. The photoionization code MAPPINGS{\sc iii} \citep{Groves04} provides a self consistent treatment of both the dust physics (photoelectric heating, dust absorption and emissivity properties, etc.) and the ISM physics, returning both line and continuum emission. The dust surrounding the cluster is considered to have contributions from three components: a population of carbonaceous grains with a power law size distribution; a population of silicate grains with the same size distribution; and a population of polycyclic aromatic hydrocarbon (PAH) molecules, whose emission is represented by a template based on IRS observations of NGC4676 and NGC7252, both interacting galaxies with strong PAH emission \citep{Groves08}. Stochastic heating is taking into account, and the maximum grain size of the distribution is $a_{\rm{max}}=0.16\: $\mum.

The radiative transfer is calculated for two physical situations. The first one considers the \ion{H}{2} region only and follows the UV photons as they traverse the wind-blown bubble, heat the dust and ionize the atomic hydrogen until the boundary of the ionization front. The second one assumes a covering photo-dissociation region (PDR) around the \hii\ region, with a hydrogen column density of $\log N(\rm{H})=22.0\: (\rm{cm}^{-2})$ (an $A_{\rm{V}}\approx1-2$ at solar metallicity). The individual model SEDs are calculated at a discrete set of ages between 0 and 10$\: $Myrs, with a resolution of 0.5$\: $Myrs, and the final integrated SED is calculated as the age-averaged energy output of the process.

In \S \ref{sec:results} we will use both the integrated models and the single \hii\ region models to interpret the observed integrated spectrum of 30 Doradus. This is equivalent to assuming two different approaches for the star formation history (SFH) of th region: an instantaneous burst of a given age, and a constant SFH over the last 10$\: $Myrs. 30 Doradus, although dominated by the single star formation event that created R136, is neither morphologically nor spectroscopically a ``single'' \hii\ region, and hence both of these simplifying assumptions should be tested to encircle the problem.

\subsection{Model Parameters}\label{sec:params}

The global parameters that represent the general assumptions of the D\&G models and that remain fixed by construction are those describing the overall geometry, the stellar IMF, the dust properties, and the PAH molecules. In the following we describe the parameters that are free to vary in the D\&G models. To reduce our parameter space and focus our analysis, in our fitting process we will keep a few of these parameters constant based on previous knowledge of the region. The free parameters are: the starburst metallicity (Z), the ISM thermal pressure ($P/k$), the cluster mass ($M_{\rm{cl}}$), the compactness ($\mathcal{C}$), the PDR fraction ($f_{\rm{PDR}}$), and the mass contained in embedded objects ($M_{\rm{emb}}$).

\subsubsection{Metallicity}

We fix the value of this parameter to $Z\approx 0.4\: Z_{\astrosun}$, which we consider a good average of several estimates using, for example, VLT observations of RR Lyrae star and Cepheid variables \citep{Gratton04} or modelling of chemical abundances in the LMC \citep{Russell92}. Metallicity variations are expected for other extragalactic starburst environments, but the well established sub-solar metallicity of the LMC helps reducing the parameter space here.

\subsubsection{ISM pressure}
This parameter describes the ambient ISM pressure that opposes the expansion of the mass-loss bubble. From a comparison between FIR line ratios in a sample of star-forming galaxies measured with the Infrared Space Observatory (ISO) and PDR models by \cite{Kaufman99}, \cite{Malhotra01} derived thermal pressures of the order of $10^5\: \rm{K}\: \rm{cm}^{-3}$. The spatial resolution achieved by ISO implies that, in most cases, this value corresponds to the thermal pressure averaged over the entire galaxy. While we acknowledge that $P_0/k = 10^5\: \rm{K}\: \rm{cm}^{-3}$ seems high for the average pressure of the LMC, we consider it a reasonable estimate near 30 Doradus, where gas densities have been boosted up by earlier star formation events. On the high pressure end, $P_0/k$ is constrained by the pressure of the ionized X-ray emitting gas inside the bubble excavated by radiation pressure near R136, which has been estimated to be of the order of $10^6\: \rm{K}\: \rm{cm}^{-3}$ \citep{Wang99}. We thus fix $P_0/k = 10^5\: \rm{K}\: \rm{cm}^{-3}$ in our models.

\subsubsection{Cluster mass}

The model SEDs scale in flux according to the total stellar mass contained in the star clusters. For an age-averaged model, averaged over the last 10$\: $Myr, the scaling relates to the total mass of stars formed during that period of time, and hence the derived mass is interpreted as a star formation rate (SFR, in $\rm{M}_{\astrosun}\: \rm{yr}^{-1}$), while for a model of a single cluster with a given age (our test case), the scaling relates to the cluster mass, $M_{\rm{cl}}$. For all cases, however, it is assumed that stochastic effects within the IMF are limited, and that the stellar population samples the full range of stellar masses.

\subsubsection{Compactness parameter}
The D\&G models introduce the compactness parameter, $\mathcal{C}$, resulting from the combination of the ISM ambient pressure $P$ and the cluster mass $M_{\rm{cl}}$. This dimensionless parameter characterizes the distribution of the ISM with respect to the ionizing stars and is based on a constant heating flux input to the stars. Intuitively, it describes how close the dust is distributed to the ionizing stars as a function of the cluster mass and hence it controls the temperature distribution of the dust and the far-IR shape of the SED. $\mathcal{C}$ is proportional to the time-averaged cluster luminosity and inversely proportional to the time-averaged square of the swept-up bubble radius. As described in D\&G, we can define the compactness as:

\begin{equation}
\label{logC}
\log \mathcal{C}=\frac{3}{5}\log\left(\frac{M_{\rm{cl}}}{M_{\astrosun}}\right)+\frac{2}{5}\log\left(\frac{P/k}{\rm{cm}^{-3}\: \rm{K}}\right)
\end{equation}

where $M_{\rm{cl}}$ is the cluster mass, $P$ is the ambient ISM pressure and $k$ is the Boltzmann constant. The pressure parameter $P/k$ relates to the ambient thermal pressure (or equivalently, the density) of the surrounding ISM.

\subsubsection{PDR fraction $f_{\rm{PDR}}$}

As mentioned above, the D\&G models explore two cases: a fully exposed \hii\ region (i.e.~the ISM ends at the ionization front), and an \hii\ region that is completely covered by the PDR in projection, with $\log N(\rm{H})=22\: (\rm{cm}^{-2})$. In reality, a star forming region will have a mix of both PDR emission and direct \hii\ emission, which we approximate by the combination of the two extreme cases, parametrized by the fraction $f_{\rm{PDR}}$: 
\begin{equation}
\label{fpdr}
F_{\nu}^{\rm{HII +PDR}}=f_{\rm{PDR}} F_{\nu}^{\rm{PDR}}+(1-f_{\rm{PDR}}) F_{\nu}^{\rm{HII}},
\end{equation}

where $F_{\nu}^{\rm{HII+PDR}}$ is the monochromatic flux arising from the star forming region, while $F_{\nu}^{\rm{PDR}}$ and $F_{\nu}^{\rm{HII}}$ correspond respectively to the fluxes calculated for the PDR-fully covered case and the \hii\ region-only case. $f_{\rm{PDR}}=0.0$ implies that there is no PDR material left around the ionized region, while $f_{\rm{PDR}}=1.0$ implies a fully PDR-covered \hii\ region.  In this fully-covered case, the PDR absorbs all of the non-ionizing UV continuum and re-radiates it at mid-infrared wavelengths. 

\subsubsection{Contribution from embedded objects}

We expect a considerable contribution from a population of massive protostars to the mid-infrared SED of \hii\ regions and starbursts, due to triggered and ongoing star formation. At the early stages of star formation, the young objects are in a protostar or Ultra-Compact \hii\ phase, deeply buried in dust envelopes. From an observational point of view, and given the age resolution of the models, these two types of objects are indistinguishable. To account for them, the models include a population of UCHIIRs \citep{Dopita06a}. In terms of the SED, these models add a component of hot dust at around 25$\: $\mum. We parametrize this contribution by scaling it to the desired mass, $M_{\rm{emb}}$.

\subsection{Attenuation by diffuse dust}

The models include an attenuation factor to account for additional absorption of UV light by foreground diffuse dust. This factor is important in the modelling of starburst galaxies, where there is significant diffuse material along the line of sight but not associated with the star-forming regions. The adopted extinction curve is derived by \cite{Fischera05} and resembles a Calzetti extinction law, which is exponential. The incoming flux is corrected for extinction as: $F=F_0 e^{-\rho*\sigma_{\rm{att}}}$, where $\rho$ is the column density of dust that gives a certain $A_{\rm{V}}$, and $\sigma_{\rm{att}}$ is the dust attenuation cross section. Based on radio continuum observations, \cite{Dickel94} find an extinction of $A_{\rm{V}}=1.1$ mag towards the 30 Doradus region. In a recent paper, \cite{Haschke11} find a reddening towards 30 Doradus of E(V-I)$= 0.43$ mag, corresponding to a similar extinction. We expect individual sources to have higher extinction values within 30 Doradus, with individual protostars having values of $A_{\rm{V}}$ up to 4.0 magnitudes. Hence for consistency we use here an average value of $A_{\rm{V}} = 2.0$.

\section{Fitting routine}
\label{sec:fitting}

We introduce here a Bayesian fitting routine for the mid-infrared SED of a starburst, either individual starbursts such as 30 Doradus, or entire starburst galaxies. This routine can be easily extended to include other wavelength ranges, and can be used for any observed spectrum that is expected  to be within the defined parameter space. We consider each model parameter as a random variable with an associated probability distribution function (PDF). Rather than just minimizing the $\chi^2$ value to find the best fitting model, we solve for the probability distribution function of each of the model parameters. 

In recent years, Bayesian analysis has been used in a number of different fields of astrophysics, where an attempt was made to reproduce a limited amount of data with multi-parameter models. Some of the applications of Bayesian methods in the determination of best fit parameters include photometric redshifts \citep{Wolf09}, observational cosmology \citep{Kilbinger10} and dusty tori around Active Galactic Nuclei (AGN) \citep{Asensio09}.

\subsection{Probability Distribution Functions}

We fit the integrated spectrum of 30 Doradus and the individual locations in Table \ref{tab:regions} using a grid of the D\&G models parametrized by the quantities described in \S \ref{sec:models}. In determining the best fit we use $\chi^2$-minimization, where the reduced $\chi^2$ is given by
\begin{equation}
\label{chisq}
\chi^2_{\rm{red}}=\sum_{i=i}^N\frac{(F_{i}-f(p_0,\lambda_i))^2} {\rm{DOF}\times \sigma_i^2},
\end{equation}

with the sum performed over all wavelength bins $\lambda_i$. The size of these bins is fixed by the wavelength resolution of the models. $F_i$ is the measured flux for each wavelength bin, $f$ is the model-predicted flux at certain wavelength $\lambda_i$ for a given set of parameters $p_0$, $\sigma_i$ is the observational error for $F_{i}$, and DOF is the number of degrees of freedom, namely the total number of wavelength bins minus the number of free parameters in the model. Minimizing $\chi^2$ gives us the best fit values for the parameters, but tells us very little about the uncertainties in the model and the parameter degeneracies. However, by exploring the $\chi^2$ surface over the range of parameters we can explore these degeneracies and the robustness of the returned parameters. 

If the errors in the parameters can be described using a Gaussian distribution, the Bayes theorem states that the probability distribution function (PDF) for a given parameter or group of parameters $(p_0)$ can be recovered from the reduced $\chi^2$ distribution:
\begin{equation}
\label{probability}
P(p_0) = \sum_{p\ne p_0} e^{-1/2\: \chi_{\rm{red}}^2},
\end{equation}

The resulting distribution, called the \textit{posterior} distribution for that parameter, is the product of the \textit{likelihood} distribution and a modulating probability distribution that includes any available a priori knowledge about the parameter, that comes from previous observations, theory, or the experimental setup. This modulating probability distribution is called the \textit{prior} distribution. We refer to the adopted prior distributions as the priors of our study.

\subsection{Model Priors}

Initially, we introduce bounded uniform priors for all parameters of the D\&G model. The bounds introduced in our priors are predominantly constrained by theory, with some constraints from observations. We use a grid of $9\times 10^5$ model outputs to cover the broad range of physical conditions in starbursts. Table \ref{tab:priors} summarizes the resulting sampling for this study, the parameter ranges and their resolutions.

\begin{deluxetable*}{cccc} 
\tablecolumns{4}
\tablecaption{Values adopted by the model parameters.}

\tablehead{
  \colhead{Parameter} &
  \colhead{Range} &
  \colhead{Resolution} &
  \colhead{Remarks} 
}
 \startdata
 Age  & $0-10\: \rm{Myr}$ & $0.5\: \rm{Myr}$ & \\
 $\log \mathcal{C}$  &  $3.5-6.5$ & $0.5$ & \\
 $f_{\rm{PDR}}$  &  $0.0-1.0$ & 0.05  & \\
 $M_{\rm{stars}}$  &  2 orders of magnitude & 0.13 dex & Adjusted to total flux density\\
 $M_{\rm{emb}}$  &  0.8 orders of magnitude & 0.05 dex & Adjusted to total flux density\\
 \enddata

\label{tab:priors}
\end{deluxetable*}

The range of ages is constrained by the typical main sequence lifetime of an early type star. For the compactness parameter $\mathcal{C}$ the limits are related to our knowledge of star-forming regions: values below $\log \mathcal{C}=3.5$ would imply very diffuse ($n \approx 10^4/T$) ISM  or stellar clusters, far lower than expected for starburst regions, while values exceeding $\log \mathcal{C}=6.5$ would lead to very compact and massive clusters. The values of stellar mass in the cluster, $M_{cl}$, and the mass contribution from embedded objects, $M_{\rm{emb}}$ are selected on a logarithmic scale depending on the total mid-infrared flux as measured in the spectra. The fraction of PDR material, $f_{\rm{PDR}}$, ranges from a completely PDR-free starburst ($f_{\rm{PDR}}=0.0$) to  a situation where the \hii\ region is completely hidden by the PDR ($f_{\rm{PDR}}=1.0$).

\subsection{Uncertainties and Models Resolution}

\subsubsection{Sources of observational error}

There are three types of errors contributing to the total uncertainty of the measured flux densities:
\begin{itemize}  
\item{The absolute flux calibration. Using model stellar atmospheres, \cite{Decin04} find that the 1$\sigma$ uncertainties on the absolute IRS flux calibration are $\approx 20\%$ for the SH and LH modules and $\approx 15\%$ for the SL and LL modules.  With regard to the modelling this error is similar to an uncertainty in the distance to the object, and affects mainly luminosity-based estimates, such as the derived SFR or stellar mass.}
\item{The relative flux calibration.  This refers to response variations within the given spectral range, often from one resolution element to the next one and is the equivalent of a ``flat field''.  From the typical differences between spectra of high signal-to-noise, taken at two different locations within the same slit, we estimated this uncertainty to be about $5\%$.  With regard to the modelling this error limits the weight that can be given to individual spectral features, and is thus a fundamental limitation to the achievable accuracy.}
\item{Systematic errors due to the specific observing conditions.  This uncertainty includes observational jitter, drifts and source (de-)centering, which may lead to a wavelength dependent change in the overall SED slope.  It also includes the amount of radiation that is external to the source of interest but picked up by the slit, e.g., from the diffuse interstellar radiation field or another nearby source.  

Flux calibration of a slit spectrum can assume that the source is a point source, and multiply by the fraction of the point source outside of the slit, which corresponds to a wavelength-dependent ``slit loss correction factor''.  Alternately one can assume that the intrinsic distribution of emission is spatially flat, so the same amount of light is lost from the slit as re-enters it from a neighbouring point on the sky. \textit{CUBISM} assumes the latter. Neither extreme is correct, and it results in a systematic flux uncertainty that scales nonlinearly with brightness. 

In addition, many adjacent spectral features, measured with the low resolution IRS modules, will be blended together.  This is most evident for the blending of the [\ion{Ne}{2}]12.81$\: $\mum\ line and the 12.7$\: $\mum\ PAH feature.  Some of these systematic uncertainties vary in time, location on the slit and wavelengths, and are extremely hard to quantify.  Hence, we do not attempt to quantify them but we need to keep these additional uncertainties in mind.}

\end{itemize}

Furthermore, many distinct spectral features, such as the nebular emission lines, contain information on the physical conditions of the ISM which is complementary to the information that can be derived from the dust continuum.  One might thus consider assigning these distinct wavelengths a larger weight (i.e., smaller error) in the fitting with respect to the more numerous continuum bins.  However, these line fluxes are difficult to model very accurately, and a larger weight combined with some mismatch between observed spectrum and model may dominate the $\chi^2$-minimization routine and lead to an incorrect local maximum in the PDF. 

One could alos consider a very sophisticated fitting procedure where each resolution element gets its unique uncertainty (i.e., weight) assigned, taking all the above mentioned error contributions.  However, we consider this approach practically impossible to reach our main goal, namely the provision of a reliable modelling procedure that yields reproducible results.  From the above arguments it is evident that any total uncertainty (which will be used as weight in the $\chi^2$ fitting) has to be larger than the error in the relative flux calibration, but will likely be smaller than the absolute flux uncertainty.  Our tests have shown that a flux uncertainty of 10\% for all IRS resolution elements leads to meaningful and robust results.  Hence, we adopt an error of 10\% per IRS resolution element.

However, the above stated values only hold for the Bright Source Limit (BSL) of the IRS, which corresponds to a S/N ratio of about 10 (\textit{IRS Instrument Handbook}).  For dim sources with S/N $< 10$ , statistical variations in the number of detected photons (i.e., shot noise) dominate the uncertainties, and our 10\% uncertainty estimate no longer holds. Other noise sources, such as noise from the detector read out  and dark current come into play, and we need a more conservative error estimate.  For all spectral resolution elements with S/N $< 10$ we use the RMS variations of the spectrum between adjacent positions in the spectral map. For each location we extract the spectra of the four nearest resolution elements.  From these five locations we calculate the average and RMS deviations for each spectral resolution element, and replace our standard 10\% uncertainty for the BSL by the RMS value for that spectral element.

\subsubsection{Data rebinning}

There is a difference between the wavelength bin size of the models and the resolution element of the Spitzer-IRS. The resolving power of the IRS ranges between 60-110, and hence a typical resolution element is of the order of 0.1$\: $\mum.  In contrast, for the models we have a wavelength step size that increases logarithmically with wavelength, and varies from 0.05$\: $\mum\ at $5\: $\mum\ to 1$\: $\mum\ at 40$\: $\mum, but includes local variations to resolve important line features. Therefore, we have re-binned the IRS spectral data to match the lower spectral resolution of the models, by averaging the fluxes of the data bins corresponding to the same model bin. The uncertainty of the resulting bins, on the other hand, are calculated as the square root of the quadratic addition of the uncertainties of the original IRS resolution elements. This propagation of error with the binning also prevents the uncertainties in the long-wavelength bins to dominate the fit.

\section{Results}
\label{sec:results}

In this section we present and discuss the results of our fitting routine applied to the integrated spectrum of 30 Doradus, and subsequently also to selected subregions within 30 Dor.  The latter have been added (see also sections 2.3 and 5.3) to probe the validity and limitations of our starburst models on regions whose spectra are dominated by one type of source (e.g., an OB cluster or a protostar).    

In the following subsection we compare the results on the integrated starburst spectrum of 30 Dor, using three different approaches for comparison: \textit{(i)} fitting all resolution elements of the entire $5-38\mu$m spectrum, \textit{(ii)} fitting the continuum bins only, i.e., excluding the fine-structure emission lines, and  \textit{(iii)}  deriving the stellar ages from the IRS high resolution lines only. We will show how sensitive the results depend on the spectral information provided. We then compare the best fit results on the integrated 30 Dor spectrum with the results on the subregions. Finally, we discuss how the results would change if we would not assume a single age burst but an age averaged model.

\subsection{Nebular lines ratios as age diagnostics.}
\label{sec:refining}

The collection of emission lines in the mid-infrared wavelength range of the IRS are sensitive diagnostics of ages of massive OB stars and to the hardness of the radiation field. They have the advantage of suffering little from obscuration and hence allow us to probe the conditions of deeply buried regions. In particular, the [\ion{Ne}{3}]$15.5\mu$m/[\ion{Ne}{2}]$12.8\mu$m and the [\ion{S}{4}]$10.5\mu$m/[\ion{S}{3}]$18.7\mu$m line ratios are good diagnostics of the ionization state of the gas, and hence provide a good constraint on the age of the ionizing cluster through measuring the hardness of the radiation field \citep[see e.g.][]{Groves08a}.

By using the ratios of the nebular emission lines, we can provide useful constraints on the ionization state and age of the central cluster, and thus break the present degeneracies. In the particular case of 30 Doradus, we have measurements of the line fluxes with both the low resolution orders (\textit{lores} hereafter) and the high resolution orders (\textit{hires} hereafter) of the IRS. The lores lines have a larger flux uncertainty, and hence a good consistency check is to compare the results we get from the routine with results derived from the hires lines. 

\begin{figure}[ht] \epsscale{1.2} \begin{center} \rotatebox{0}{\plotone{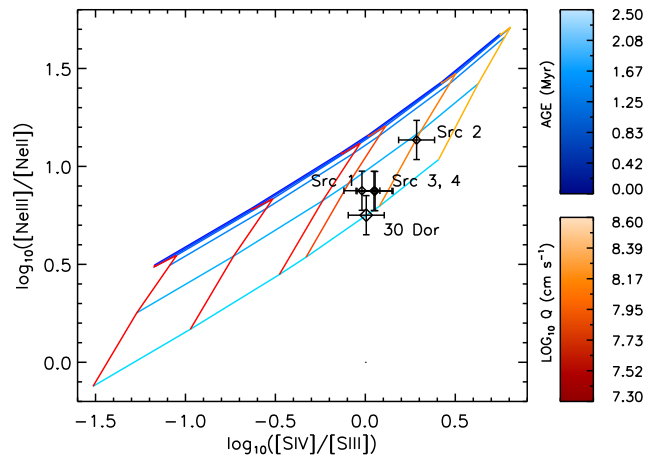}} \end{center} \caption{\label{fig:levesque_age} Measured mid-infrared nebular line ratios overplotted on a grid of starburst models from \citet{Levesque10} that include a single \ion{H}{2} region only. The parameters of the model are the age of the stellar cluster, and the ionization parameter, Q (the ratio of ionizing photon density to gas density), as labelled on the color-bars to the right.} \end{figure}

We use the hires nebular line fluxes presented in \cite{Lebouteiller08} as a reference for the estimation of cluster age and ionization parameter and compare the results with what we obtain from the SED fitting for the four individual sources of Table \ref{tab:regions}. These line ratios are plotted in Fig.~\ref{fig:levesque_age} superimposed on to a grid of models by \citet{Levesque10} \citep[created using the ITERA program of][]{Groves10}. These are essentially the same as the D\&G models, and use both the Starburst99 and MAPPINGS{\sc iii} codes with similar assumptions about the gas. However, the \citet{Levesque10} models use a much simpler geometry (namely plane-parallel instead of spherical), demonstrating much more clearly how the degeneracy between the hardness of the radiation field (i.e.~stellar cluster age) and the ionization parameter, Q (the ratio of the ionizing photon density to gas density), is broken using four strong mid-infrared emission lines. The dependence of the line ratios on these two parameters has also been note by \cite{Morisset04}. Our comparison between the measured line ratios and the predictions from the Levesque models indicate ages between 2.0-2.5$\: $Myr for all four positions.

The sample of sources in \cite{Lebouteiller08} includes five more locations in the 30 Doradus region, apart from our four selected sources. Assuming that the measured line fluxes in these sources are representative of the overall conditions in the cluster, we estimate the line ratios for the whole region from the luminosity-weighted average line fluxes: $\log_{10}$[\ion{Ne}{3}]/[\ion{Ne}{2}]$_{\rm{30Dor}} = 0.75$ and $\log_{10}$[\ion{S}{4}]/[\ion{S}{3}]$_{\rm{30Dor}} = 0.005$. We also plot this average value in Fig.~\ref{fig:levesque_age}. The result indicates an age of 2.5$\: \rm{Myr}$. 

In Table \ref{tab:posages} we list the ages derived from the Lebouteiller et al.~hires line ratios as compared to the ages derived from the SED fitting in two cases: \textit{(a)} fitting the emission lines and \textit{(b)} excluding the emission lines. The reason to perform the fit using the continuum only is two-fold. First, we want to check the consistency of the results for our individual sources, which can not be treated as isolated \hii\ regions, since their ionization states are affected by other external sources. Second, as we have already stated, while the line ratios can be reasonably estimated by the models, the lores equivalent widths of the lines are predicted with a lower degree of accuracy by the D\&G models.

\begin{deluxetable*}{ccccc} 
\tablecolumns{5}
\tablecaption{Ages of individual sources}

\tablehead{
  \colhead{Source} &
  \colhead{SED fit, with lines (Myrs)}  &
  \colhead{SED continuum (Myrs)}  &
  \colhead{Line ratios (Myrs)}  
}
 \startdata
 1   &  7.5  &  2.0   & 2.0-2.5 \\
 2  & 10.0  &  2.5     & 1.5-2.0 \\
 3  &   8.0  &  2.5     & 2.0-2.5  \\
 4  &    0.5   &  2.0    & 2.0-2.5  \\ 
30 Dor &  3.0 &   5.0  & 2.5\\
\enddata

\label{tab:posages}
\end{deluxetable*}

The results show that for the individual sources, the ages derived from the continuum-only fit are consistent with the high resolution measurements from Fig.~\ref{fig:levesque_age} and with the independent measurements of the overall age of the region. On the other hand, including the unresolved lines in the fit for these individual sources leads to age estimates which are in disagreement with all the other methods, with a tendency to overestimate the ages. 

The lowres line ratios imply age estimates that are not significantly different from the hires results and hence the mentioned disagreement should be interpreted in terms of the limitations of the integrated \hii\ region D\&G models to reproduce the continuum and the line emission for \emph{individual} sources.  Nonetheless, for the integrated spectrum, the derived age from the emission line fit is consistent with the high resolution measurements and with the literature, as expected for a self-contained region for objects of which class the models were intended.

Based on this results, for the integrated spectrum of 30 Dor we run the fitting routine including the emission lines. For the individual sources, however, we do not attempt to fit the low resolution lines and fit only the continuum. Also for the individual sources, to include the information contained in the high resolution line measurements, we modify the prior probability distribution for the ages from those listed in Table \ref{tab:priors} to use a Gaussian distribution centered a 2.5$\: $Myrs with a dispersion of 1.5$\: $Myrs. This suppresses weights solution with older ages down, further constraining the parameters.

\subsection{Integrated Spectrum}

\subsubsection{Best fit}
\label{sec:best_fit_int}

\begin{figure}[ht] \epsscale{0.95} \begin{center} \rotatebox{90}{\plotone{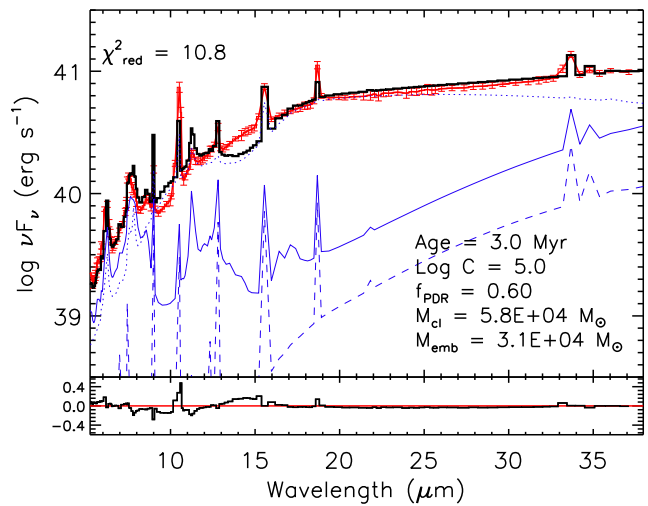}} \end{center} \caption{\label{fig:best_fit_int} Best fit to the integrated IRS spectrum of 30 Doradus. \emph{Red}: Observed spectrum rebinned to the model resolution with error bars for each bin. \emph{Black}: best fit SED. \emph{Dashed blue}: ``Naked'' \hii\ region contribution. \emph{Solid blue}: PDR and obscured \hii\ region contribution. \emph{Dotted blue}: Embedded object contribution. The best fit values and reduced $\chi^2$ are indicated. Residuals are shown in the lower panel, in the same logarithmic units.} \end{figure}

We show the resulting best fit from our code to the integrated spectrum of 30 Doradus in Fig.~\ref{fig:best_fit_int}, with the residuals of the model fit to the observations shown in the lower panel. The quality of the fit is remarkable, with most of the spectral features in the mid-infrared spectral range successfully reproduced. This is a significant improvement from broad band photometry SED fitting, where only a few data points were fitted to constrain an equal number of parameters.

The individual contributions from the unobscured \hii\ region, the PDR, and embedded populations are explicitly plotted in Fig.~\ref{fig:best_fit_int}. The residuals in the bottom panel indicate that the model fits the observations within the uncertainties for most of the IRS wavelength range, but underestimate the fluxes near 15$\: $\mum. This feature is most likely due to an overabundance of small silicate grains within the assumed dust model, and is dominated by the embedded star model. 

The sulphur lines at 10.5$\: $\mum\ and 18.3$\: $\mum\ appear as underestimated by our best fit model. This could be possibly due to abundance and/or pressure variations \citep[see][]{Dopita06c}, and we can only argue here that for the assumed abundances and ISM pressure, the fit in Fig.~\ref{fig:best_fit_int} represents the best case in the prediction of line ratios. The mid-infrared continuum is dominated by the embedded population, especially for $\lambda > 10\: $\mum\, with the PDR contributing mainly to the PAH fluxes and the continuum slope at the long wavelength end of the spectral range. The \hii\ region and PDR are responsible for most of the emission lines.

\subsubsection{Interpretation of the results for the integrated spectrum}

\begin{figure*}[ht]
  \centering
  \subfigure[Cluster mass]{
\includegraphics[scale=0.23,angle=90]{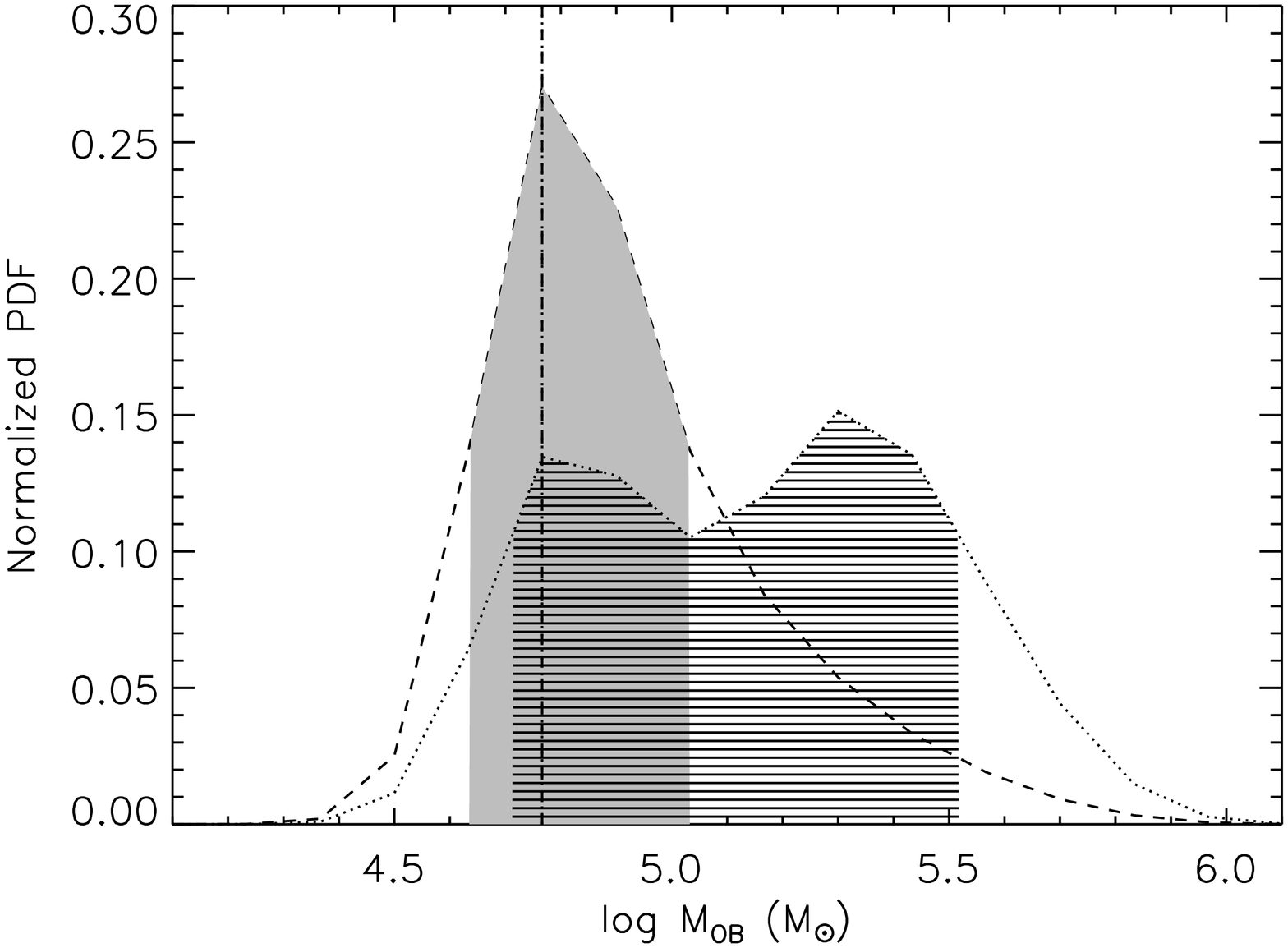}

}
  \subfigure[Mass of embedded objects]{
\includegraphics[scale=0.23,angle=90]{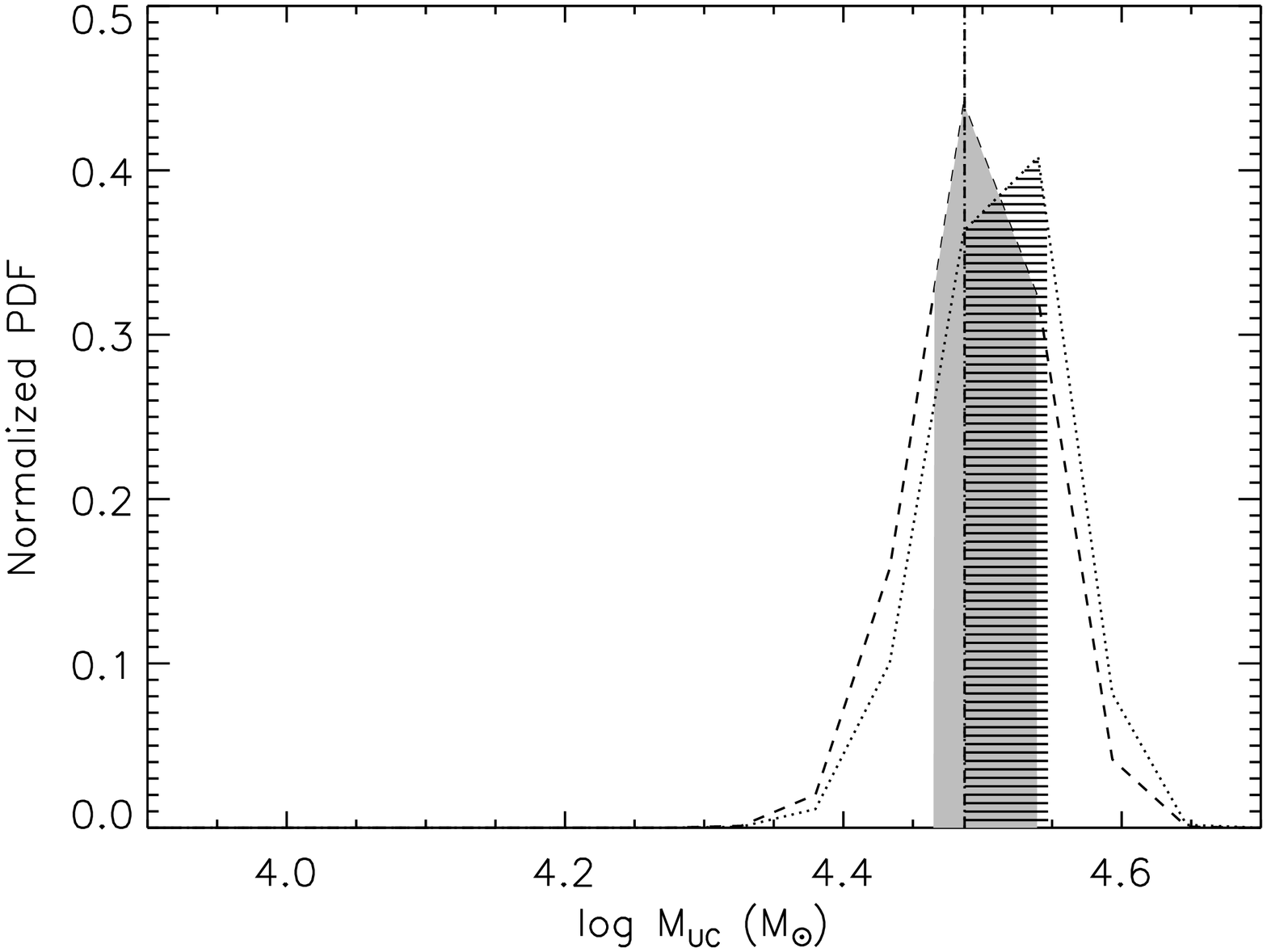}

}
\subfigure[Age]{
\includegraphics[scale=0.23,angle=90]{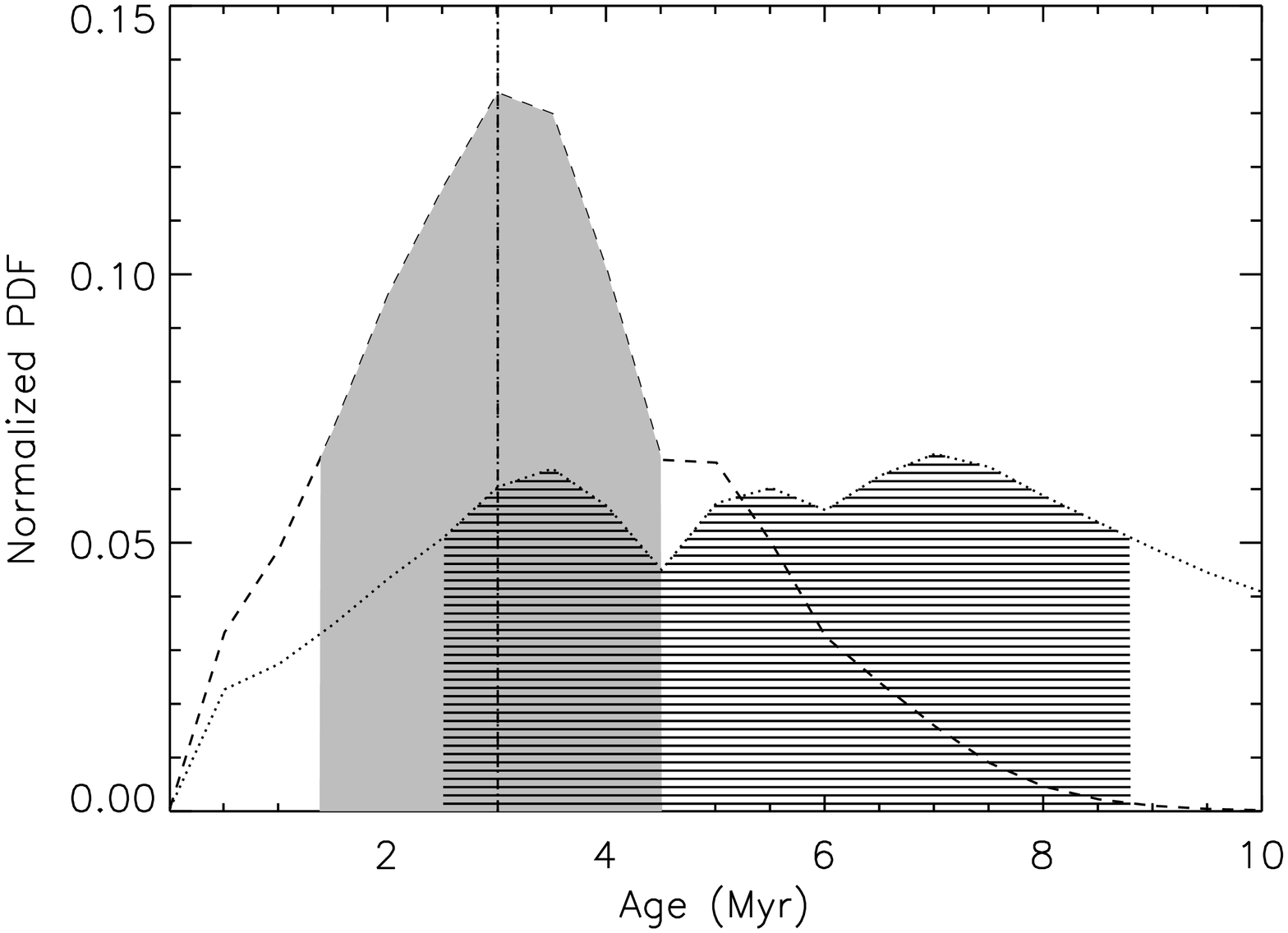}

}
\subfigure[PDR fraction]{
\includegraphics[scale=0.23,angle=90]{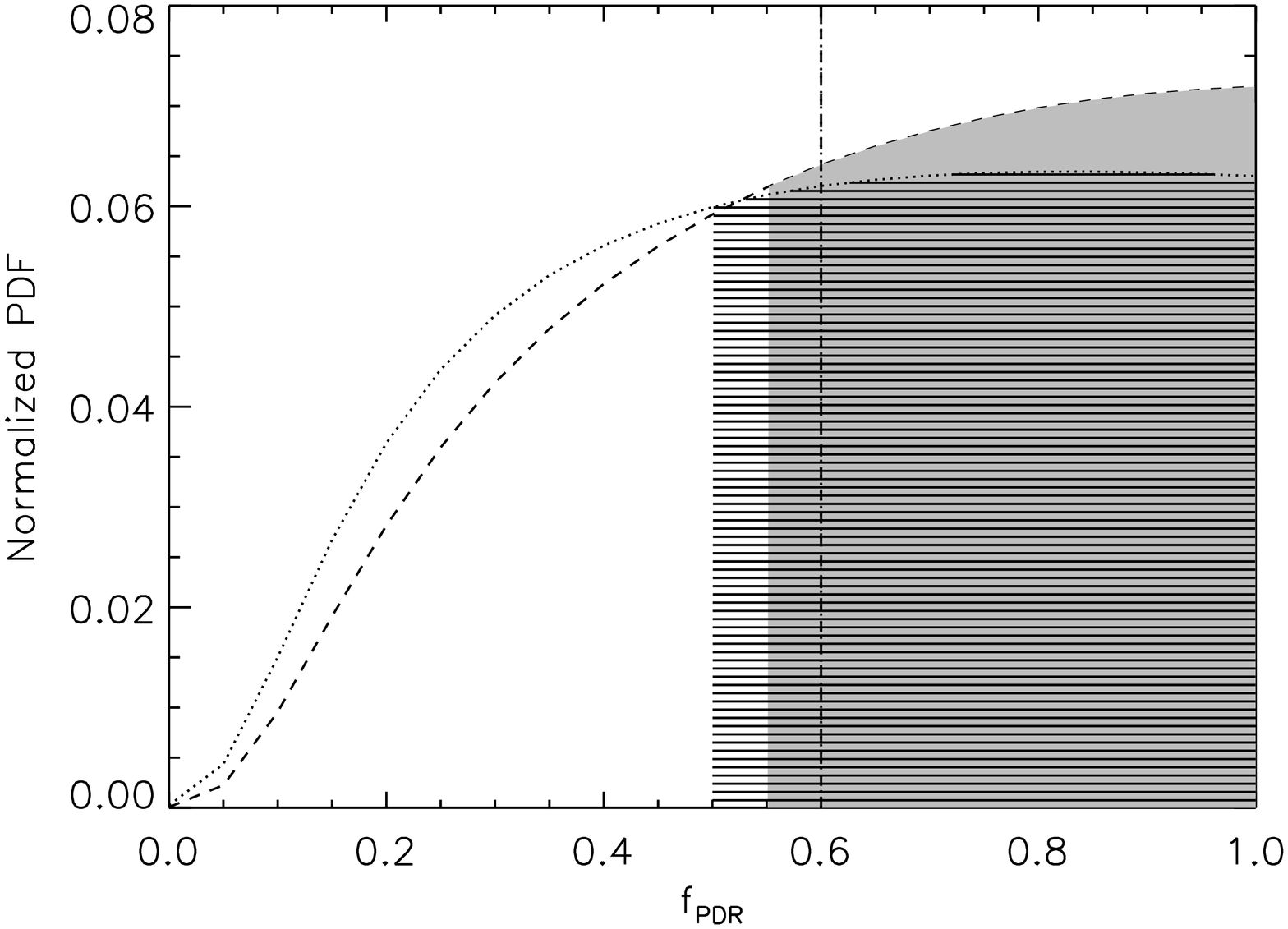}

}
\subfigure[Compactness]{
\includegraphics[scale=0.23,angle=90]{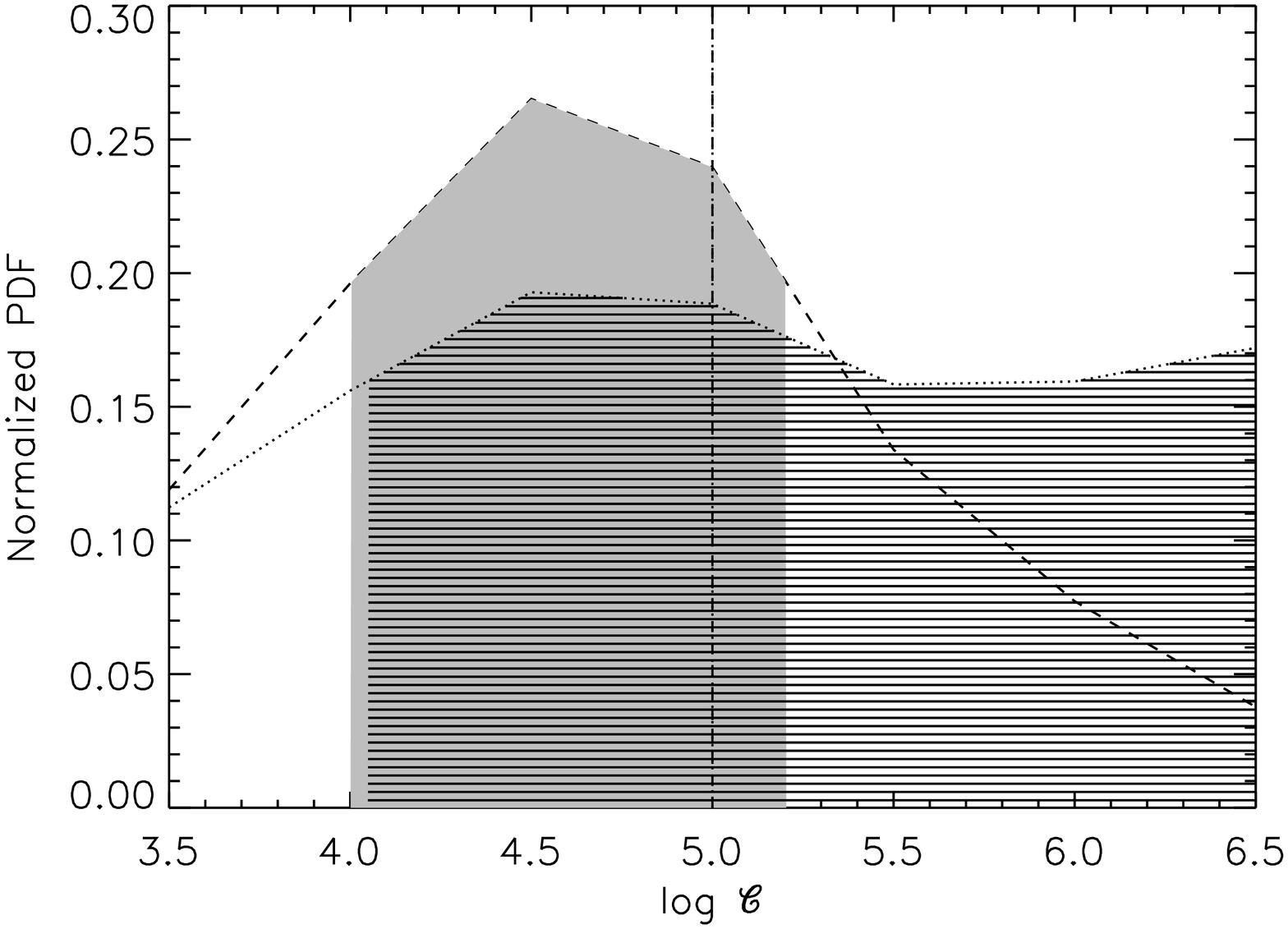}

}
  \caption{Probability distribution functions of the model parameters when fitted to the integrated spectrum of 30 Doradus, for two cases. \textit{Dotted line}: uniform priors described in Table \ref{tab:priors}. \textit{Dashed line}: Modified prior distribution for age, as defined in \S \ref{sec:refining}. The shaded areas correspond to the 1-$\sigma$ integrated probabilities, while the vertical lines indicate the best-fit values.}

\label{fig:PDFs_30Dor}
\end{figure*}

The normalized PDFs for the model parameters are shown in Fig.~\ref{fig:PDFs_30Dor}, for the priors as listed in Table \ref{tab:priors} (dotted lines), and for the modified probability distribution of ages described earlier (dashed lines). The best fit values marked by vertical lines and the 1$\sigma$ uncertainties indicated by the horizontal line pattern. We list the best-fit values, with the uncertainties corresponding to each case, in Table \ref{tab:bestfit}. 

\begin{figure*}[ht]
  \centering
  \subfigure[]{
\includegraphics[scale=0.23,angle=90]{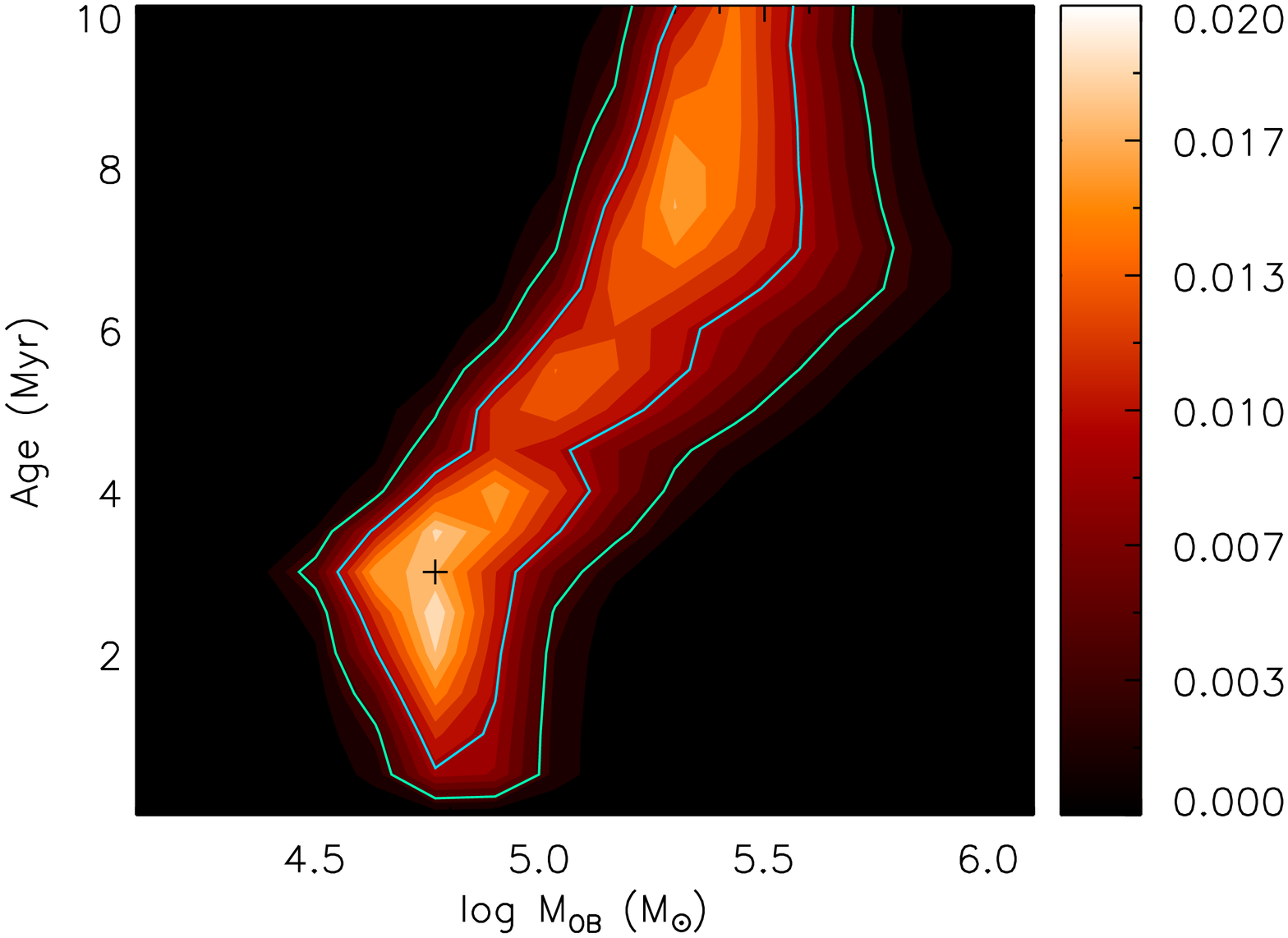}

}
  \subfigure[]{
\includegraphics[scale=0.23,angle=90]{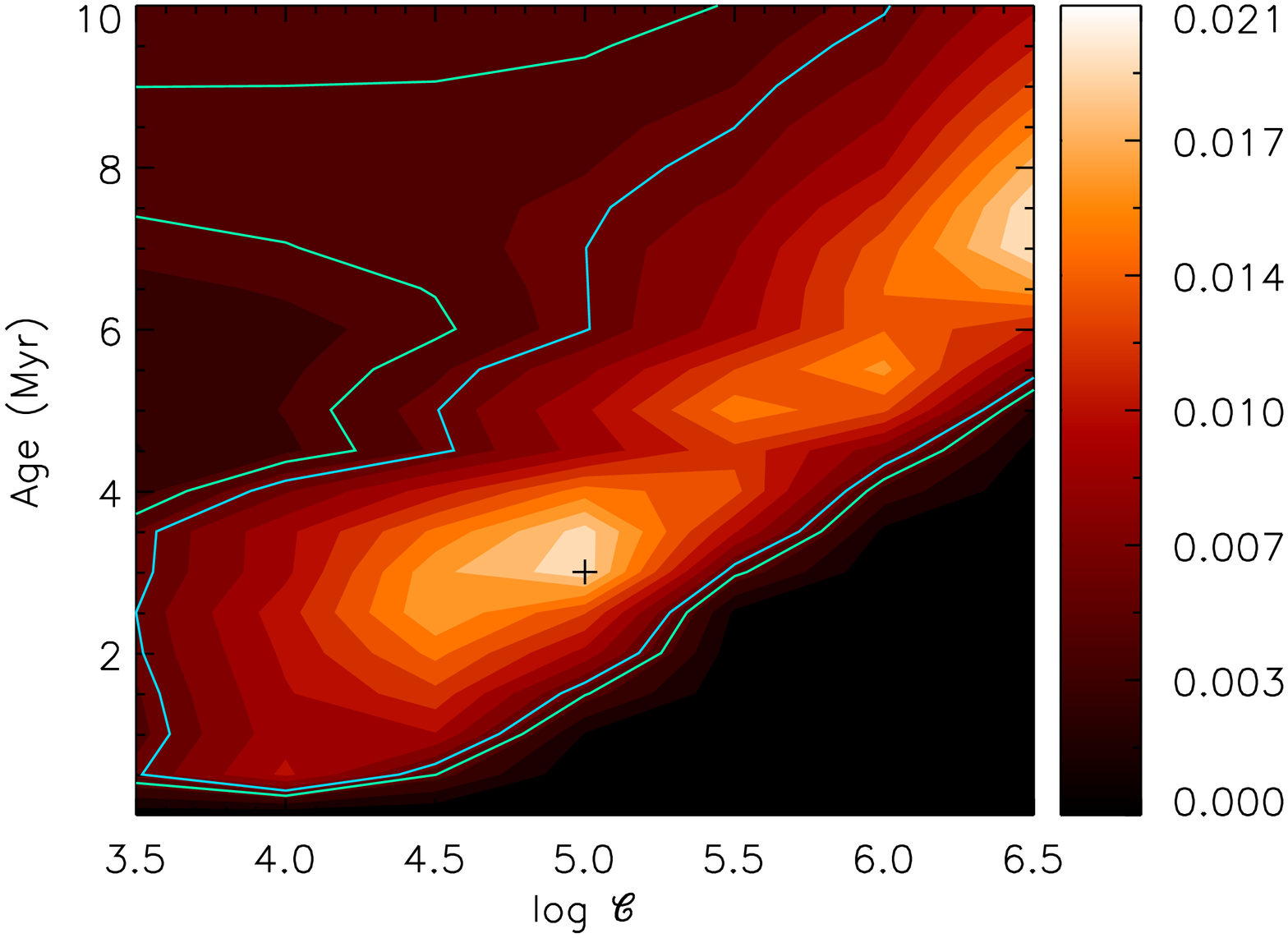}

}
\subfigure[]{
\includegraphics[scale=0.23,angle=90]{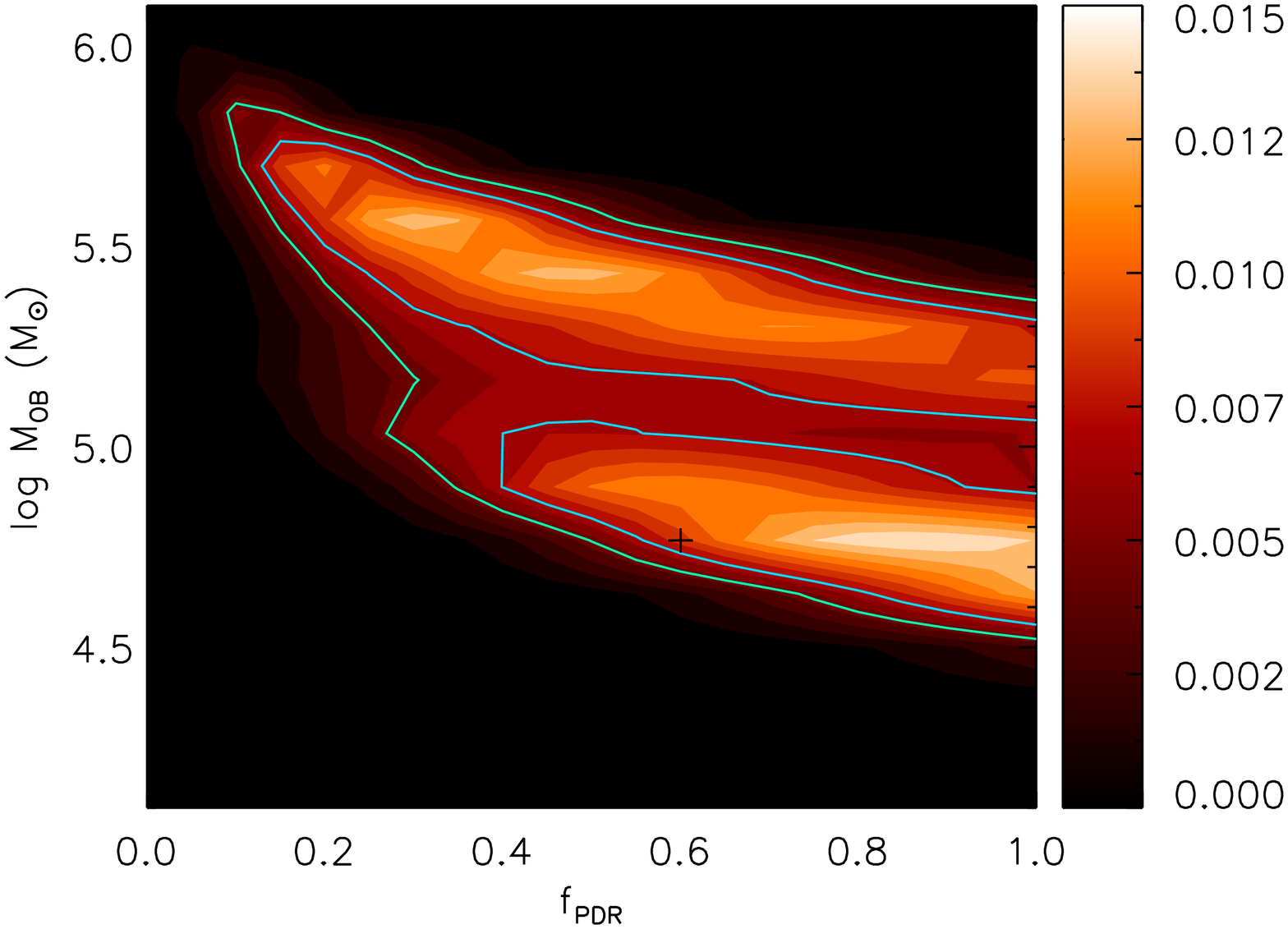}

}
 
  \caption{Two-dimensional PDFs for selected pairs of parameters showing the model degeneracies. The color code indicates normalized probability. The cross symbols mark the best-fit values while the color contours indicate the 1-$\sigma$ (blue) and 90\% (green) confidence levels.}

\label{fig:2D_PDFs_int}

\end{figure*}

From the dark-shaded and dotted PDFs in Fig.~\ref{fig:PDFs_30Dor} it is evident that several of the parameters appear to be very broad or even unconstrained. The reason for this can be clearly seen when we plot 2D PDFs for selected pairs of parameters in Fig.~\ref{fig:2D_PDFs_int} (i.e. collapsing the $\chi^2$ space down to two parameters). These show degeneracies between certain model parameters, indicating that, at least in the IRS wavelength range, these parameters affect the SED shape in a similar way. If one or both of these parameters can be constrained using other information, such as from other wavelengths, the 1D PDFs should become narrower, and the parameters better constrained.

\begin{deluxetable*}{cccc}
\tablecolumns{4}
\tablecaption{Best fit to the integrated spectrum of 30 Doradus}

\tablehead{
  \colhead{Parameter} &
  \colhead{Age-unconstrained} &
  \colhead{Age-constrained} &
  \colhead{Literature} 
}
 \startdata
  $Age$ (Myr)                                                 & $3.0_{-0.5}^{+5.5}$     &   $3.0_{-1.5}^{+1.5}$   &     $\approx 3$ \citep{Hunter95}      \\ [3pt]
  $\log \mathcal{C}$                                    & $5.0_{-1.0}^{+1.5}$     &     $5.0_{-1.0}^{+0.2}$     &                                                             \\ [3pt]
 $f_{\rm{PDR}}$                                                &  $0.6_{-0.1}^{+0.4}$    &    $0.6_{-0.1}^{+0.4}$       &                                                                  \\ [3pt]
 $\log M_{\rm{cl}}$ ($\rm{M}_{\astrosun}$)       &   $4.8_{-0.1}^{+0.7}$    &    $4.8_{-0.2}^{+0.2}$      &      $4.7$ for R136 only \citep{Hunter95}   \\ [3pt]
 $\log M_{\rm{emb}}$ ($\rm{M}_{\astrosun}$)  & $4.47_{-0.02}^{+0.08}$     &   $4.47_{-0.02}^{+0.08}$  &                                                                   \\ [3pt]
 \enddata

\label{tab:bestfit}
\end{deluxetable*}

In order to understand these degeneracies we need to look carefully at the 2D probability maps and link the resulting distributions to the effect that each parameter has on the spectrum. There is an age-compactness-cluster mass degeneracy revealed by two different set of parameters that provide a good fit to the observed SED. The first panel of Fig.~\ref{fig:2D_PDFs_int} clearly shows the resulting two-peak distribution on the probability distribution for the total cluster mass-age subspace. These two parameters, as well as compactness, have a similar effect on the mid-infrared continuum as they vary across the grid: they scale the continuum flux by certain multiplicative factor. 

The age-mass component of this degeneracy is not surprising: as compared to a cluster of certain mass and age, a less massive cluster is dimmer at Spitzer wavelengths, but the same holds true for an older cluster. IRS continuum fitting alone is incapable of distinguishing between these two parameters, as can be seen from the two-peak distribution. However, we have more information contained in the nebular lines. In particular, older clusters show less nebular emission, as the ionizing radiation strongly decreases with age. Including the lines in the fit breaks the age-mass degeneracy and enables us to select, between the two possible solutions, the one that best reproduces the measured line ratios. The best fit in Fig.~\ref{fig:best_fit_int} corresponds to this best solution.

The two peaks of the compactness-age degeneracy are clearly seen in the PDF maps, as shown in the second panel of Fig.~\ref{fig:2D_PDFs_int}. This degeneracy implies that both a young cluster with small compactness or an old cluster with high compactness lead to similar fits of the observed spectra, provided that the cluster mass also adjusts. This is shown in the central panel of Fig.~\ref{fig:2D_PDFs_int}. In this case, the emission lines are less sensitive to the variations of the two parameters together, since both the age of the cluster and the compactness affect the line ratios. However, once the age has been determined from the line ratios, the compactness probability distribution also shrinks and selects only one of the possible solutions for compactness.

\begin{figure*}[ht]
  \centering
  \subfigure[]{
\includegraphics[scale=0.23,angle=90]{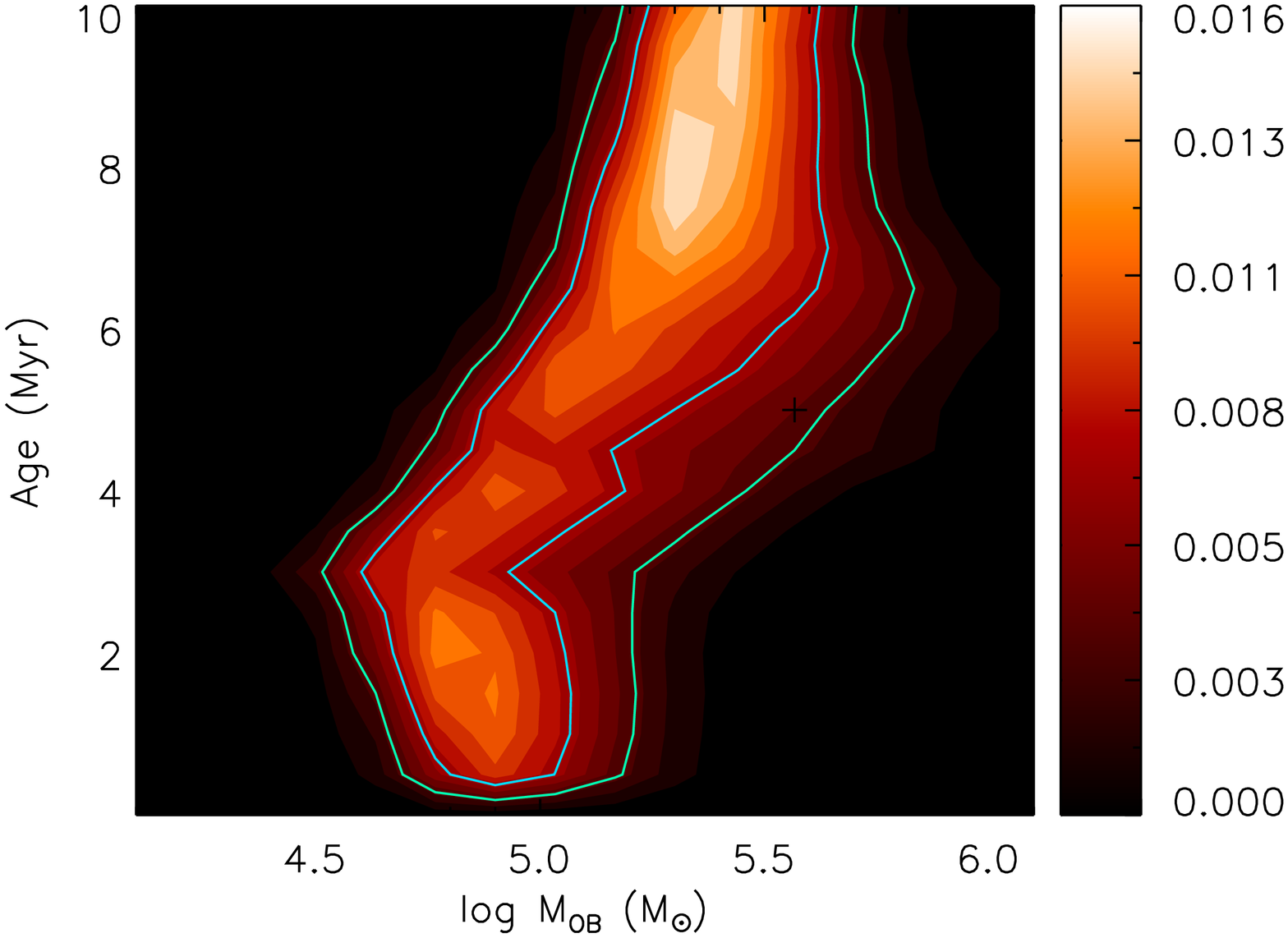}

}
  \subfigure[]{
\includegraphics[scale=0.23,angle=90]{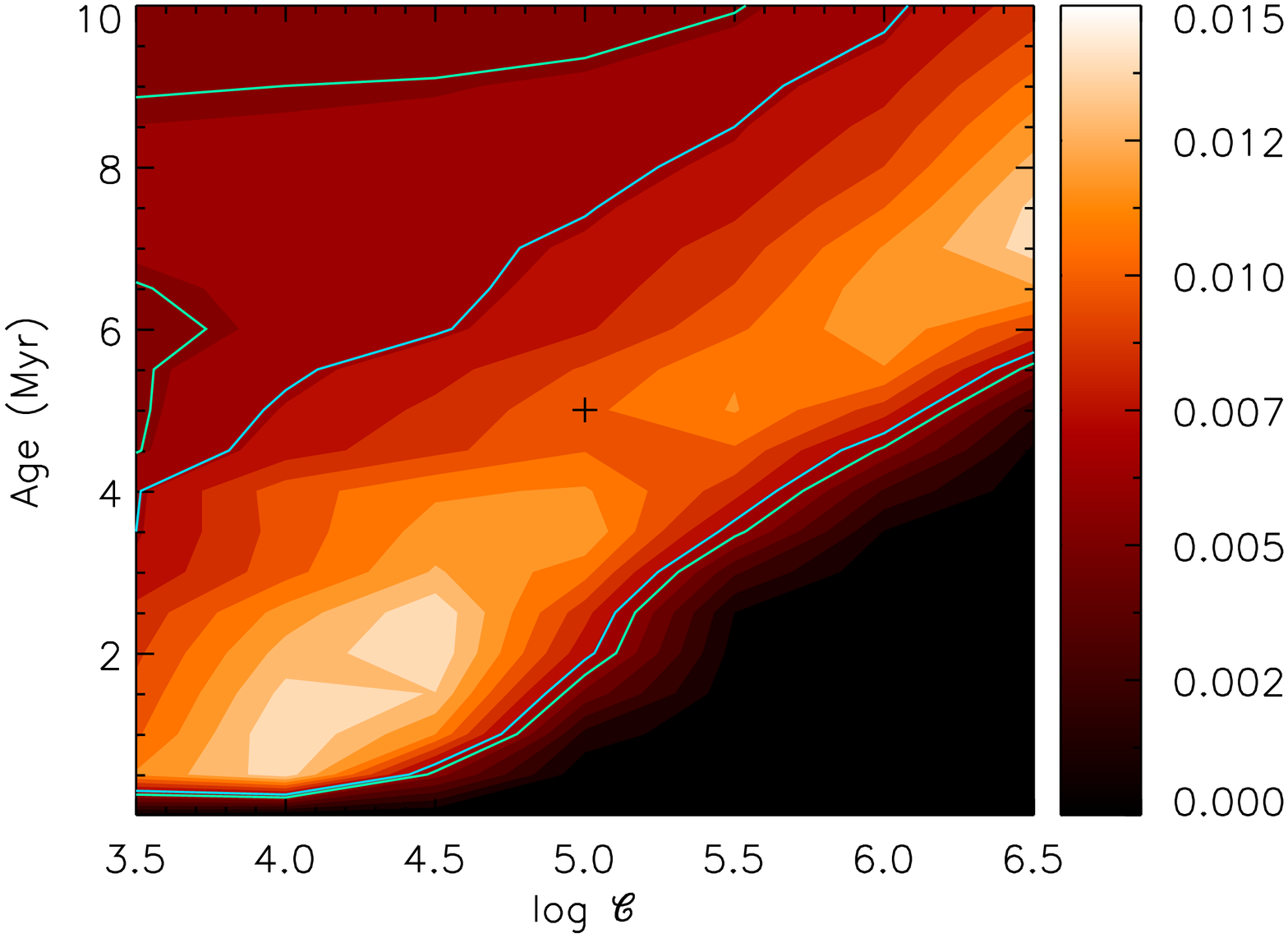}

}
\subfigure[]{
\includegraphics[scale=0.23,angle=90]{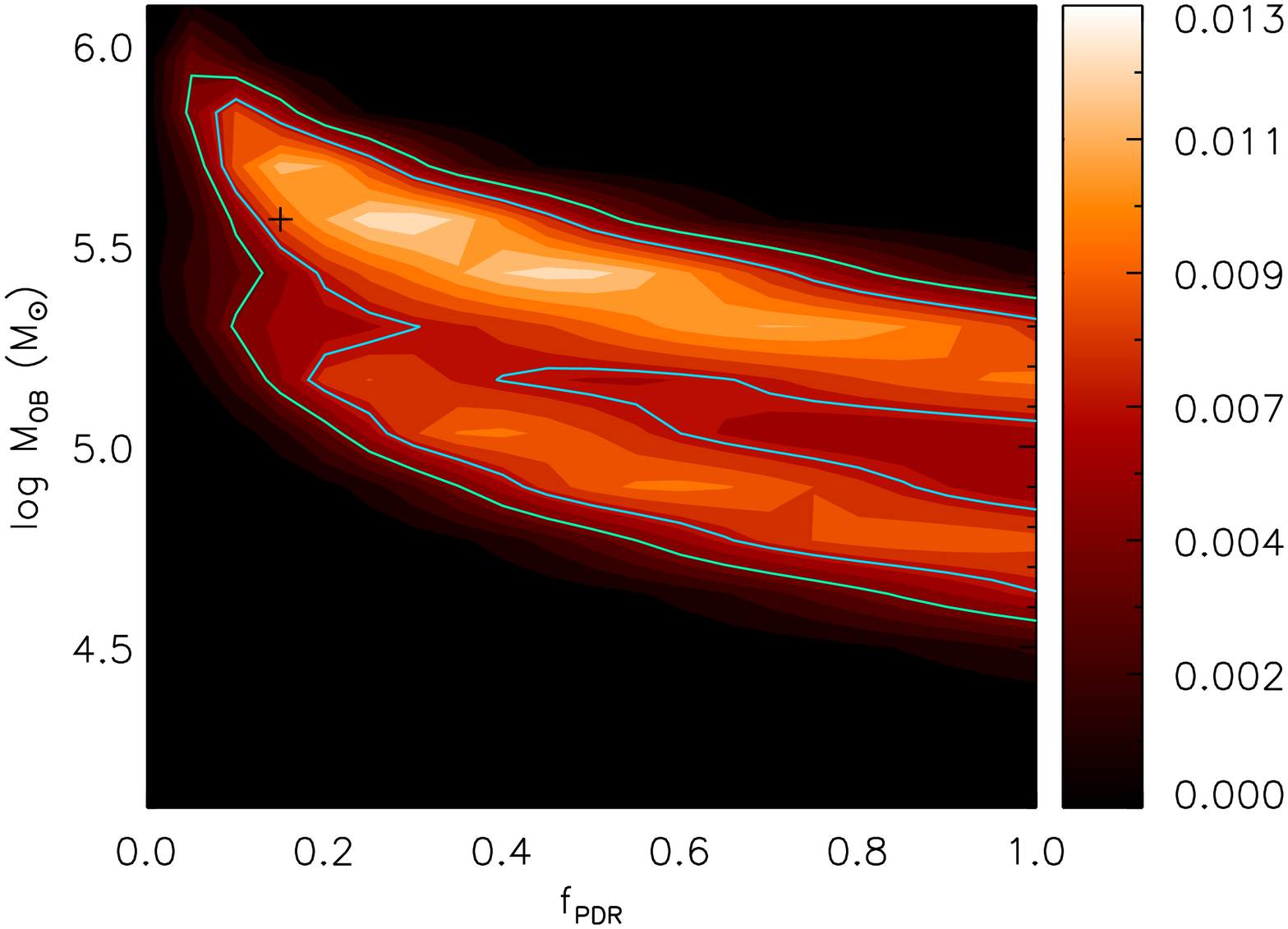}

}

  \caption{Two-dimensional PDFs for selected pairs of parameters, when only the continuum has been fitted. The color code indicates normalized probability. The cross symbols mark the best-fit values while the color contours indicate the 1-$\sigma$ (blue) and 90\% (green) confidence levels.}

\label{fig:pdfs_no_lines}
\end{figure*}

Panel (c) of Fig.~\ref{fig:2D_PDFs_int} shows a degeneracy between cluster mass and PDR fraction. The two strips correspond to the two peaks of the age-mass-compactness degeneracy, while the smooth diagonal variation corresponds to the PDR fraction-cluster mass degeneracy. This degeneracy arises from the fact that the PDR region that covers the \hii\ region contributes mostly PAH emission, but also adds thermal dust continuum that in the  models scales up with the PDR fraction. The emission lines are not of great help in breaking this degeneracy, since their relative fluxes are almost insensitive to variations in PDR content and total mass. In this particular point, thus, we can only do better if we include data from other wavelengths.

To assess the importance of the line ratios in the constraining of the parameters and the break of the age-compactness-mass degeneracy, in Fig.~\ref{fig:pdfs_no_lines} we plot the PDFs for the same parameter pairs, but this time after only the continuum has been fitted. A quick comparison between the two cases reveals that the inclusion of the lines not only selects one of the two degenerate peaks, but also helps the best fit values to converge towards the absolute maximum of the PDF. This is particularly evident for the mass-age degeneracy, where the continuum-only fit favors an older age solution, while the inclusion of the lines shifts the probability maximum to an age that is in agreement with independently measured values.

\subsection{Individual sources}

As noted before, any complex starbursting system is likely to cover a wide range of object types and physical conditions, from individual protostars and luminous UCHIIRs to OB clusters, loose stellar associations, PDRs and the diffuse ISM.  To investigate the range of conditions for which our starburst models still yield accurate results, we have chosen four subregions that probe these ``extreme'' cases where one of these components is expected to dominate the mid-infrared spectrum, based on detection of, for example, infrared excess or X-ray emission.  These sources are the OB cluster R136 with little dust obscuration, an \ion{H}{2} region which shows high extinction along the line of sight (source 3), and two compact objects, which are luminous protostellar candidates (sources 2 and 4).

We fit the continuum spectra of those subregions and include the emission line ratios as a modified probability of the ages, as described in \S \ref{sec:refining}. While we do not expect to get a very good fit on these types of sources with the general starburst models, we want to verify that the crucial parameters are qualitatively still constrained within reasonable limits, according to the respective physical condition probed by the individual sources.

\subsubsection{Best fits to the spectra of individual sources}
We have seen that in the case of the integrated spectrum, fitting the emission lines along with the continuum greatly helps in breaking the model degeneracies. However, as discussed in \S \ref{sec:refining} and summarized in Table \ref{tab:posages}, this is not the case for the individual sources, where the fitting of unresolved lines leads to age estimates which are in disagreement with the continuum-only fit and with the line ratio analysis, for reasons that are described in \S \ref{sec:refining}. Hence, we do not include the emission lines in the SED fitting of the individual sources. Furthermore, we modify the age prior to include only ages that are consistent with the high-resolution line ratios, within the uncertainty limits set by the comparison of the line ratios and the Levesque models. In each case, instead of the uniform prior distribution of probability for the age, we use a Gaussian PDF centered at 2.0$\: $Myr with an uncertainty of 1.5$\: $Myr. 

\begin{figure*}[ht]
  \centering
  \subfigure[Source 1 - R136]{
\includegraphics[scale=0.47,angle=90]{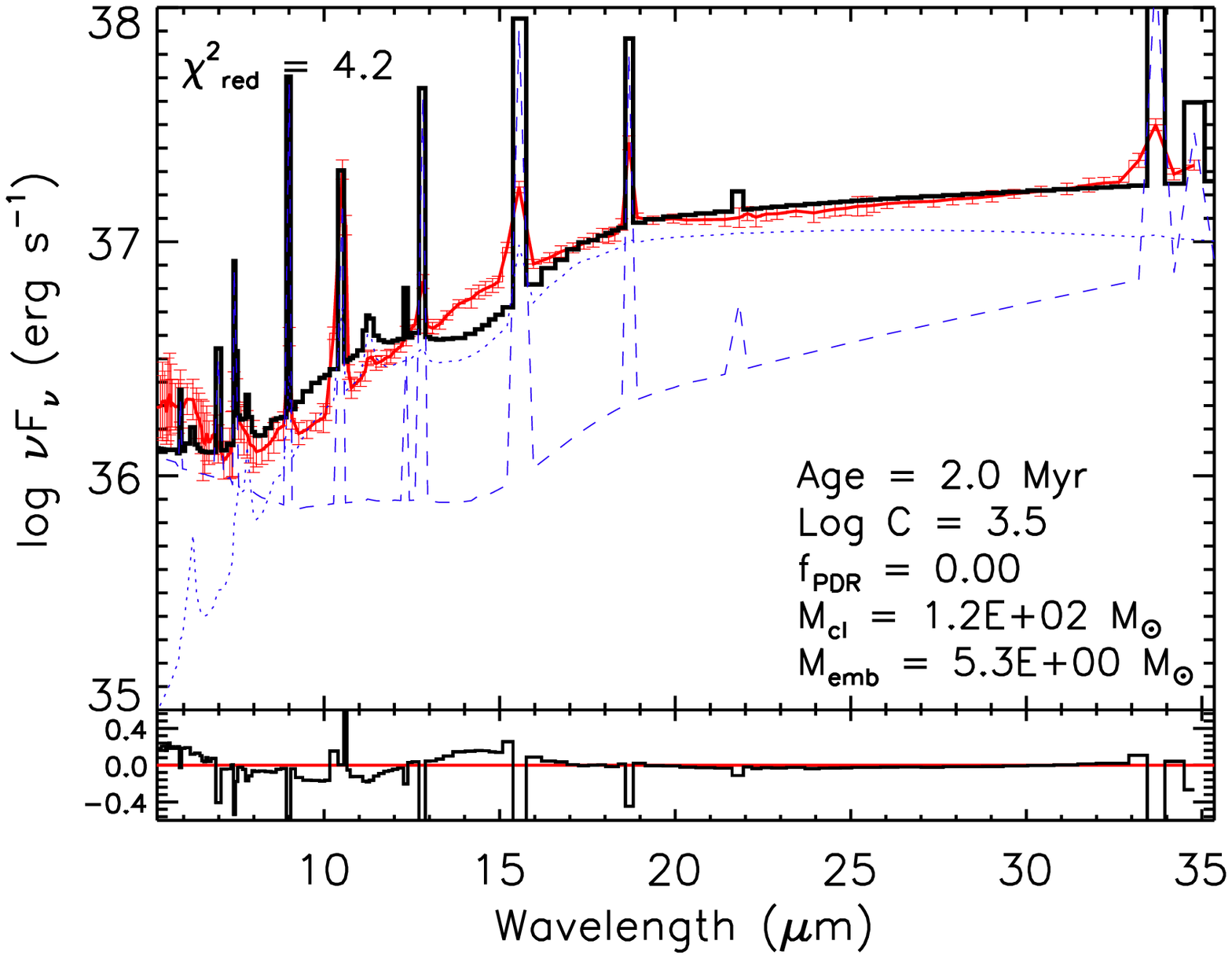}

}
  \subfigure[Source 2 - Protostar candidate]{
\includegraphics[scale=0.47,angle=90]{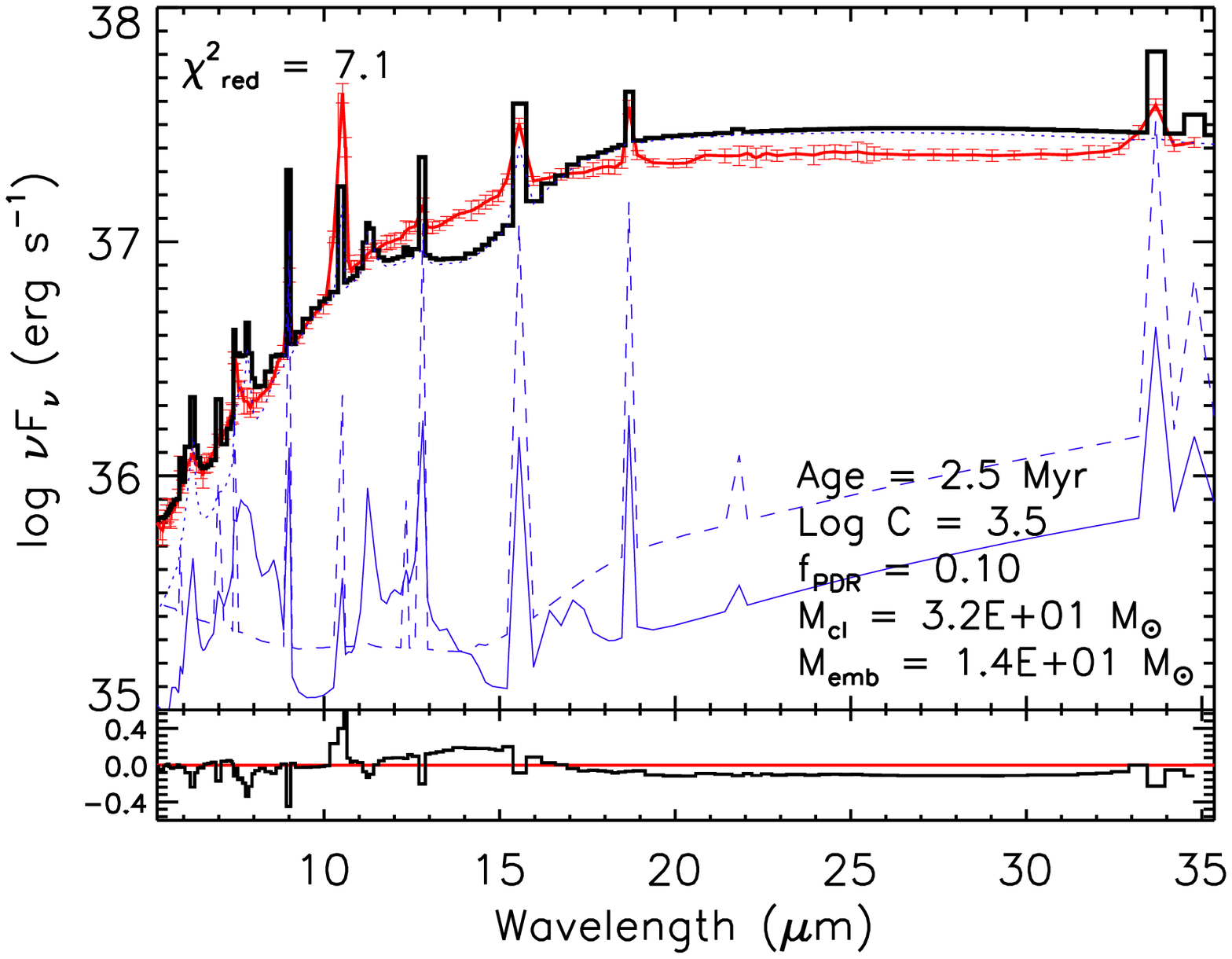}

}
\subfigure[Source 3 - Highly extincted source]{
\includegraphics[scale=0.47,angle=90]{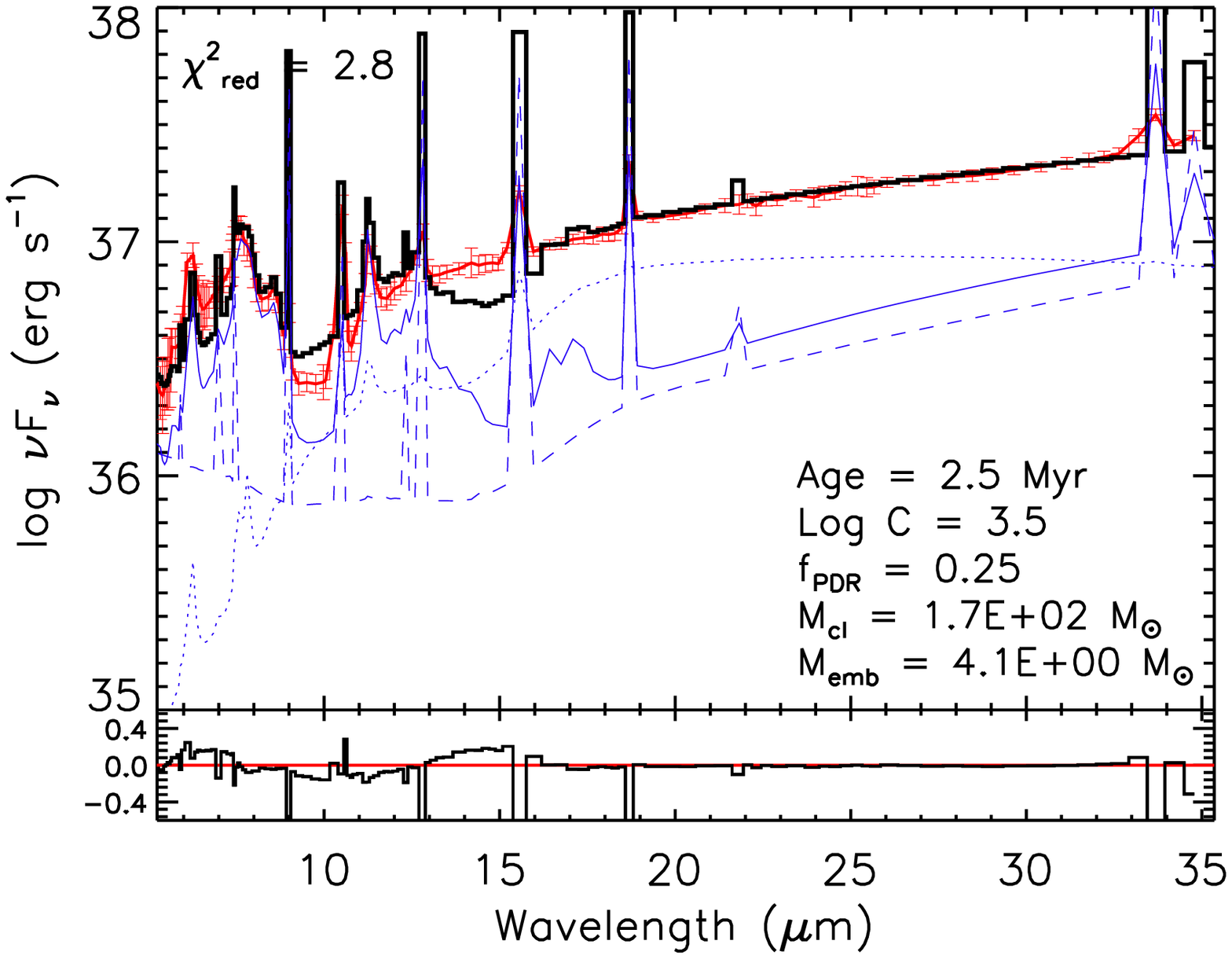}

}
\subfigure[Source 4 - Protostar candidate]{
\includegraphics[scale=0.47,angle=90]{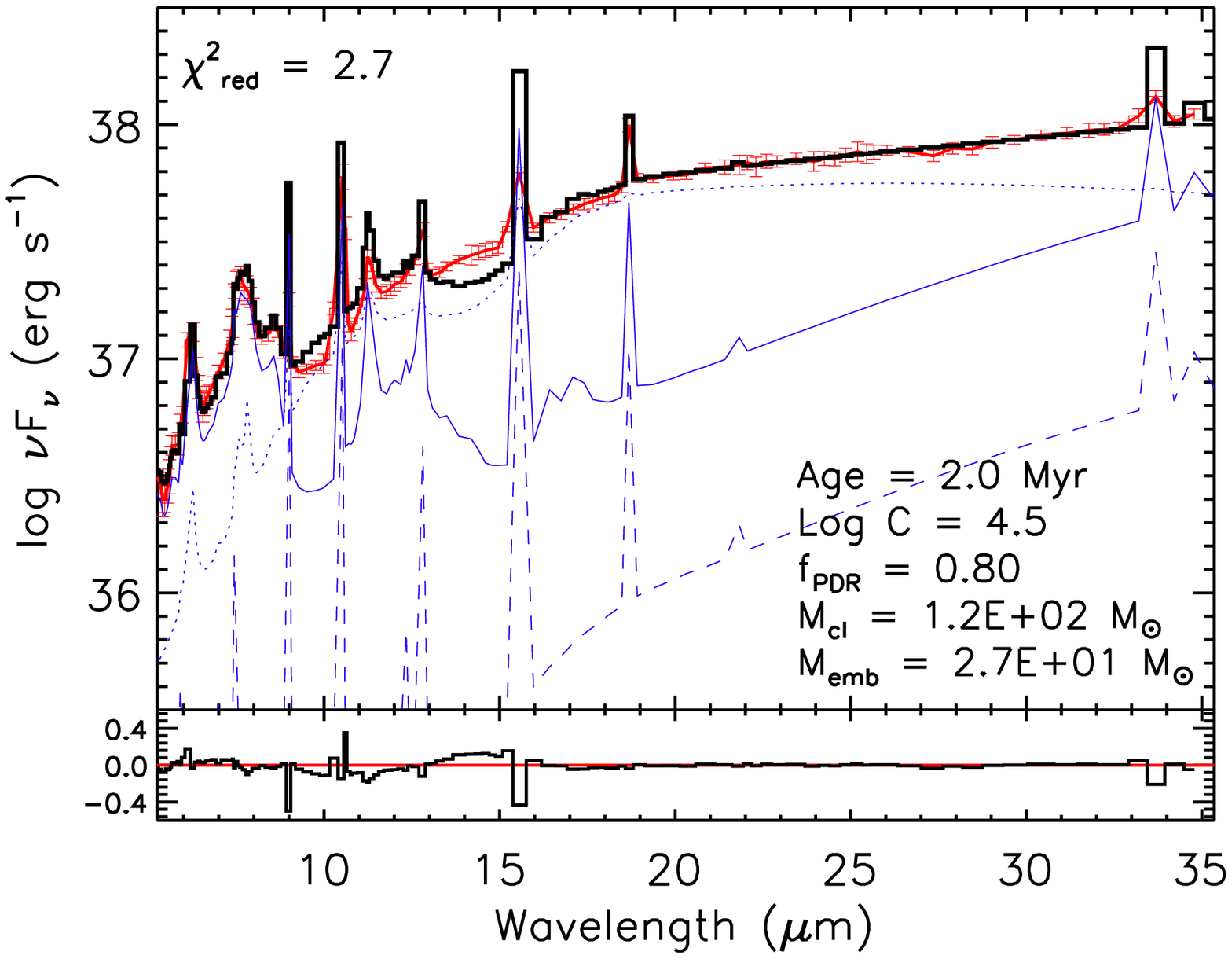}

}

 \caption{Best fit models for the individual positions, excluding the emission lines from the fit and with the age distribution constrained by the line ratio analysis. The color code is the same as for Fig.~\ref{fig:best_fit_int}.}

\label{fig:best_fit_pos}

\end{figure*}

The resulting best-fit SEDs are shown in Fig.~\ref{fig:best_fit_pos}. The resulting best fit parameters and 1-$\sigma$ ranges derived from the PDFs are shown in Table \ref{tab:bestfitpos}. 

\begin{deluxetable}{ccccc}{b}

\tablecolumns{5}
\tablecaption{Best fit parameters for individual positions in 30 Dor}

\tablehead{
  \colhead{} &
  \colhead{Source 1} &
  \colhead{Source 2} &
  \colhead{Source 3} &
  \colhead{Source 4} 
}
 \startdata
  $t$ (Myr)                                                    &  $2.0_{-1.0}^{+2.0}$          &    $2.5_{-1.5}^{+2.0}$        &    $2.5_{-1.0}^{+2.5}$      &  $2.0_{-1.0}^{+2.5}$     \\ [3pt]
 $\log \mathcal{C}$                                     &  $3.5_{-0.0}^{+1.0}$         &    $3.5_{-0.0}^{+0.5}$        &   $3.5_{-0.0}^{+0.5}$       &  $4.5_{-0.5}^{+0.5}$      \\ [3pt]
 $f_{\rm{PDR}}$                                                &  $0.00_{-0.00}^{+0.40}$    &    $0.10_{-0.05}^{+0.50}$   &   $0.25_{-0.00}^{+0.70}$  &  $0.80_{-0.35}^{+0.20}$       \\ [3pt]
 $\log M_{\rm{cl}}$ ($\rm{M}_{\astrosun}$)       &  $2.1_{-1.1}^{+0.0}$          &    $1.5_{-0.6}^{+0.0}$        &   $2.2_{-0.5}^{+0.2}$       &  $2.1_{-0.2}^{+0.4}$       \\ [3pt]
 $\log M_{\rm{emb}}$ ($\rm{M}_{\astrosun}$)     &  $0.73_{-0.03}^{+0.12}$      &  $1.13_{-0.03}^{+0.07}$     &   $0.61_{-0.06}^{+0.14}$  & $1.43_{-0.08}^{+0.12}$   \\ [3pt]
 \enddata

\label{tab:bestfitpos}
\end{deluxetable}

\subsubsection{Interpretation of the results for individual sources}

All of our spectra show significant flux densities in the 10$\: $\mum\ range.  This continuum emission is indicative of hot dust at $T\approx 300$\,K, which is typically associated with protostars, but not exclusively.  It may also include emission from dust close to slightly more evolved stars, as well as hot dust in between stars of a cluster, and dense clumps in the \ion{H}{2} region that cannot be modelled by the simple uniform \ion{H}{2} region model of D\&G.  We account for all these contributions by what we have called the embedded component.  Fig. \ref{fig:best_fit_int} shows that this component dominates the emission of 30 Doradus at mid-infrared wavelengths.  

In fact, our attempts to fit the integrated spectrum of 30 Doradus without including this embedded component have proven unsuccessful, and hence, it is one of our main results that this component of ``embedded objects'' is necessary to fit the observed spectra for $\lambda > 10\: $\mum.  Qualitatively, this interpretation looks quite plausible: the two positions that coincide with the location of YSO candidates (sources 2 and 4) have the higher relative mass contribution from the embedded component, with 30\% and 20\% of the total mass contained in embedded objects, respectively. The corresponding contributions from embedded mass in R136 and the highly extincted source are 5\% and 2\%, respectively (Table \ref{tab:bestfitpos}). However, we need to keep two issues in mind:

First, while we associate this component with a recently formed star, strictly speaking it is not entire due to protostars and UCHIRs, for the reasons given above.  Hot dust in starbursts may also be found in other environments and, hence, the amount of ``embedded objects'' derived from our fits can only be considered as an {\em upper limit} on the amount of protostars and UCHIRs. 

Second, the contribution of this embedded component to the total emission at $\lambda > 10\: $\mum\ may be surprisingly high but is not unreasonable.  This is illustrated in Figs. \ref{fig:best_fit_pos}(a) and \ref{fig:best_fit_pos}(b), which show the fits to source 1 (R136, the main cluster) and source 2 (a protostar, or group of protostars). While $M_{\rm{cl}}$ (stars + \hii\ region + PDR ensemble) is at least one order of magnitude larger in the case of R136 than in the embedded region, the contribution of the embedded component is much higher for the protostar in comparison, and so are the observed flux densities over most of the IRS spectral range.  While R136 contributes most of the stellar mass, much less massive components described as ``embedded objects'' contribute the majority of the mid-IR flux.  

In other words, the fact that the integrated flux in 30 Doradus is dominated by this embedded component does not imply that there is a similar mass contribution from this embedded component. In fact, our derived mass contribution of embedded objects to the stellar mass of 30 Doradus is about 35\%. Furthermore, despite the uncertainty in the nature of the embedded component, we will show in section 5.4 that the star formation rate in 30 Dor derived from our modelling approach does {\em not} overestimate the ``true'' star formation rate as derived from integrated panchromatic SEDs that include the far-IR. Additionally, the application of our routine to starburst galaxies shows that this component of embedded objects does not dominate the mid-infrared emission for these galaxies, as it does for 30 Doradus.

Our results in Table \ref{tab:bestfitpos} indicate that source 4 has a higher compactness as compared to the other individual sources. High values of $\log \mathcal{C}$ are expected in compact starbursts with high surface brightness, where dust is in close proximity to intense UV fields. Source 4 is a very bright and compact source of [\ion{S}{4}]10.5$\: $\mum, indicating the presence of highly ionized gas probably near a hard UV source, and, as pointed out in \S \ref{sec:locations}, is also a bright X-ray source that has even been considered as a supernova remnant candidate. No other location in our spectral map shares these characteristics as an individual source. On the other hand, low compactness is derived for sources where the simultaneous presence of bright stars and dust can be inferred, as it is the case of sources 1, 2 and 3.

We conclude that, even though our models are not intended to model these individual sources, we can nonetheless learn from the compactness parameter, as defined in \S \ref{sec:models_brent}, by comparing the results of the routine applied to them. The routine is capable of constraining the proximity of luminous sources and hot dust. We have derived a relatively high compactness for 30 Doradus itself with $\log \mathcal{C}=5.0$, which is consistent with the luminous cluster in its center surrounded by nearby ridges of dust.

A comparison of the values for $f_{\rm{PDR}}$ in Table \ref{tab:bestfitpos} with the spectra in Fig.~\ref{fig:all_spectra} indicates that, as expected from the construction of the models, a high covering fraction is generally associated with strong PAH features. We show here that source 2, where a YSO has been identified via infrared excess, has little associated PAH emission, whereas source 4, which also shows infrared excess and has been associated with a YSO, shows significant PAH emission. The presence of PAH emission in the line of sight towards embedded objects might depend on the evolutionary stage of the YSO, i.e., on the optical thickness of the envelope. If UV photons from a young, massive protostar manage to escape the embedded region and create a PDR around the YSO, then we expect to detect PAH emission. In some cases, like source 3, where the deep silicate feature typical from highly embedded objects is accompanied by high line ratios indicative of an ionizing source, the PAH emission can even dominate over the embedded component. No significant PDR emission is inferred from the fit to R136. This is consistent with the expected absence or low abundance of PDR material very close to the ionizing cluster. 

There are two important caveats that we must consider in the interpretation of the PDR covering fraction. First, we have not included here emission from diffuse dust, not associated with the starburst (i.e.~ heated by the interstellar radiation field); this adds both cold dust and PAH emission. Second, there exists a degeneracy between the covering fraction of PDR material and the cluster mass in the mid-infrared (right panel of Fig.~\ref{fig:2D_PDFs_int}), due to the fact that PDR regions not only add PAH emission, but also continuum emission at longer wavelengths. This translates in the apparent mismatch between the PDFs and the best fit values for these two parameters in Table \ref{tab:bestfitpos}.

\subsection{Age averaged case}
\label{sec:age_average}

The degeneracy found for the integrated spectrum of 30 Doradus that leads to the two possible solutions of an old, massive cluster or a young cluster with a smaller mass is indicative that not even 30 Doradus can be considered as a single coeval stellar population. As described in the introduction, a number of spectroscopically identified populations have been identified in 30 Doradus, and the fact that the continuum is compatible with two different sets of parameters leads us to the conclusion that a simple, single-age approach might be inaccurate even for our benchmark \hii\ region. Thus we also carry out the SED fitting using an age-averaged model.

\begin{figure}[ht] \epsscale{0.95} \begin{center} \rotatebox{90}{\plotone{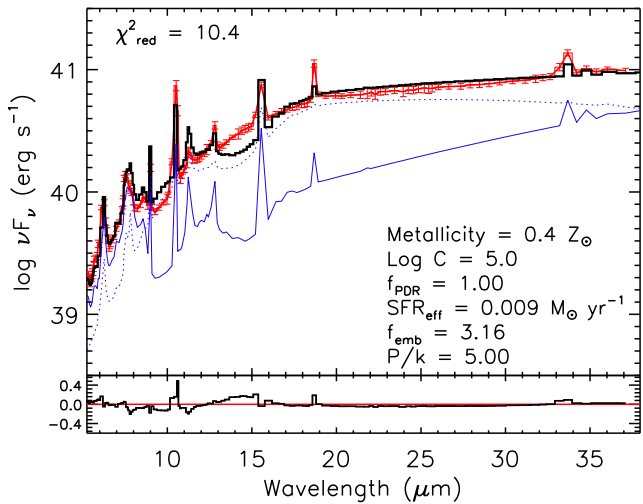}} \end{center} \caption{\label{fig:bf_age_average} Best fit to the integrated IRS spectrum of 30 Doradus using the age-average model. The color code is the same as in Fig.~\ref{fig:best_fit_int}.} \end{figure}

We have fitted the integrated spectrum of 30 Dor using the age-average model described in \S \ref{sec:models_brent}, where the SED is integrated over the 0--10$\: \rm{Myr}$ lifetime of the ionizing stars. In this case, the cluster age is no longer a free model parameter, and we can interpret the absolute flux scaling as the SFR instead of a single cluster mass. We keep the ISM pressure ($P/k$) and the metallicity ($Z$) fixed to the same values as for the single age case. The best fit parameters we calculate in this way must be interpreted as average values over the time span covered by the models. Fig.~\ref{fig:bf_age_average} shows the resulting best fit to the data with the age-average model, and Table \ref{tab:age_average} lists the best-fit parameters with the associated uncertainties derived from the PDFs. 

\begin{deluxetable}{cc}
\tablecolumns{2}
\tablecaption{Best fit to the integrated spectrum of 30 Doradus with age-average model}

\tablehead{
  \colhead{Parameter} &
  \colhead{Best fit} 
}
 \startdata
  
  $\log \mathcal{C}$                                                         & $5.0_{-1.0}^{+0.0}$                \\
 $f_{\rm{PDR}}$                                                                     &  $1.00_{-0.45}^{+0.00}$           \\
 $\log \rm{SFR_{\rm{eff}}}$ ($\rm{M}_{\astrosun}\: yr^{-1}$)    & $-2.05_{-0.05}^{+0.35}$          \\
 $f_{\rm{emb}}$                                                                    &  $3.20_{-0.40}^{0.64}$              \\
 
 \enddata

\label{tab:age_average}
\end{deluxetable}

Here we use the parameter $f_{\rm{emb}}$ instead of $M_{\rm{emb}}$. It refers to the  ratio of mass contained in embedded objects to the mass of main sequence stars for objects younger than 10$\: $Myr, and hence it is related to the amount of currently ongoing star formation. In other words, $f_{\rm{emb}}$ gives the fraction of embedded/UCHII luminosity-weighted contribution that we have to add to the SED to fit the observed spectrum. If $f_{\rm{emb}}=0.0$, there is no current star formation happening, whereas if $f_{\rm{emb}}=1.0$, half of the massive stars formed over the last million year are still in a embedded state. Since this contribution is integrated over a period of 1$\: $Myr only, adding embedded objects also implies that the average SFR has to be modified according to:

\begin{equation}
\label{eq:SFR}
{\rm{SFR}}_{\rm{eff}}=\rm{SFR}+\frac{f_{\rm{emb}}}{f_{\Delta t}}\rm{SFR}
\end{equation}

where ${\rm{SFR}}_{\rm{eff}}$ is the effective SFR that accounts for the additional population of embedded objects, and $f_{\Delta t}$ is the ratio of the total time over which the starburst has been modelled (10$\: $Myr) and the estimated duration of the embedded phase (1$\: $Myr). For our best fit case, we get $\rm{SFR}=0.007\: \rm{M}_{\astrosun}\: \rm{yr}^{-1}$ and $f_{\rm{emb}}=3.2$, which implies that among the stars younger than 1$\: $Myr, there are about three times more embedded objects than main sequence objects. Hence, the effective SFR is ${\rm{SFR}}_{\rm{eff}}=\rm{SFR}+0.32\times \rm{SFR} = 0.009\: \rm{M}_{\astrosun}\: \rm{yr}^{-1}$.

The LMC has a SFR of 0.1$\: \rm{M}_{\astrosun}\: \rm{yr}^{-1}$ \citep{Whitney08}. Keeping in mind that this is only a lower limit estimate, given the incompleteness of any YSO catalogue, our result implies that between 3\% and 10\% of the star forming activity of the LMC takes place in the 30 Doradus region. To see how this compares to estimates of the SFR in 30 Doradus from single photometric measurements, we compare the IRAS flux at 25$\: $\mum\ for the entire LMC to the 64$\: $pc$\times$63$\: $pc area from which we have extracted the spectrum of 30 Dor. The total flux density from the LMC at $25\: $\mum\ is $7520\pm 1100\: $Jy \citep{Israel10}. From our integrated spectrum (Fig.~\ref{fig:all_spectra}), we derive a monochromatic flux density of $1739\pm 174$ Jy for the same wavelength, which corresponds to 24\% of the total LMC flux density. Assuming this wavelength directly traces star formation, it suggests that our value is close, but may be underestimated by a factor of two. 

The compactness and fraction of PDR results are consistent in both the single age case and the age average case.

\section{Summary and conclusions}
\label{sec:discusion}

Significant progress in our understanding of starburst systems has been made over the past decades, both on the observational and theoretical side.  A huge amount of spectral data on star forming regions and starburst galaxies has been collected with ISO, Spitzer and the Herschel Space Telescopes, complemented by a considerable library of SED models that predict the energy output of starbursts as a function of wavelength.

However, SED fitting of starburst has mainly focused on maximum likelihood methods, which generally overlook degeneracies between physical parameters and lead to results that are not unique.  Furthermore, these ad hoc approaches often depend on some hidden assumptions that make the results reproducible. In this paper, we presented a routine to fit the SEDs of starbursts based on the models proposed in the series of papers \citet{Dopita05}, \citet{Dopita06b}, \citet{Dopita06c} and \citet{Groves08}.  We verified the accuracy and limitations of our approach by comparison between the model fit results and the known properties of the well-studied, prototypical giant \hii\ region 30 Doradus.  Our main findings are:

\begin{itemize}

\item Our modelling procedure is able to fit a broad range of continuum slopes, PAH intensities, and emission lines.  Although we have only used the mid-infrared spectra for the calibration, the method can be easily expanded to other wavelength ranges.

\item We have verified the validity of our approach by comparison with the well studied 30 Doradus region.  The derived physical parameters, such as cluster mass, cluster age, ISM pressure and PDR content, are in good agreement with the known properties of this nearby starburst.

\item We have provided a detailed study of the model degeneracies in the mid-infrared window of the spectrum, and have shown that the best fit values to the continuum shape are driven by a triple luminosity-age-compactness degeneracy that, in general, leads to multiple ``best fits''.

\item The inclusion of emission lines in the analysis breaks this degeneracy. It is expected that the addition of other wavelength ranges would further constrain the model parameters. In particular, the precise location of the peak of the dust emission in the far-infrared is crucial to constrain the compactness parameter. Herschel spectroscopy, as well as MIPS and PACS photometry, play an important role here.

\item We provided meaningful results to the model-defined compactness parameter $\mathcal{C}$, introduced by \citet{Groves08}, and linked them to the proximity of ionizing sources and hot dust.

\item We have shown that modelling the SED of a typical starburst region requires a component of heavily embedded objects (massive YSOs and UCHIIRs) which dominate the mid-infrared continuum slope.  The derived mass fraction of this embedded component can be interpreted as an upper limit to the amount of current star formation, since there are other dust heating mechanisms not included in the models.

\item We found a degeneracy between the total stellar mass and the relative amount of PDR material, $f_{\rm{PDR}}$, as both will contribute to the dust continuum. This degeneracy may lead to uncertain mass estimates and can only be resolved with additional data at longer wavelengths, e.g. from Herschel.

\item Generally, two critical assumptions in all starburst models are the age and duration of the burst.  Our ``local'' template 30 Doradus nicely illustrates the typical complexity of a starburst with both, the presence of a luminous, coeval cluster (R136), and strong evidence for continuous star formation across the region. Hence, we have also used an age-average model of continuous star formation for comparison.  This model delivers values for compactness and PDR contribution that are consistent with those derived from the single age models.  For 30 Doradus we derive a contribution of approximately 10\% to the total SFR of the LMC.
\end{itemize}

Now with a robust and well-tested modelling and fitting routine in hand, we plan to apply this approach to more distant giant \hii\ regions and starburst galaxies.  The lack of spatially resolved data on e.g., more distant ULIRGs and sub-millimeter galaxies requires reliable and well calibrated models to derive the physical conditions in these starbursts.  The novel fitting procedure presented in this paper constitute the next step in starburst modelling and puts such studies on solid grounds.


\begin{thebibliography}{widest-label} 
%
%

\bibitem[Allen et al.(2004)]{Allen04} Allen, L.~E., et al.\ 2004, \apjs, 154, 363

\bibitem[Andersen et al.(2009)]{Andersen09} Andersen, M., Zinnecker, H., Moneti, A., McCaughrean, M.~J., Brandl, B., Brandner, W., Meylan, G., \& Hunter, D.\ 2009, \apj, 707, 1347

\bibitem[Asensio Ramos \& Ramos Almeida(2009)]{Asensio09} Asensio Ramos, A., \& Ramos Almeida, C.\ 2009, \apj, 696, 2075

\bibitem[Beir{\~a}o et al.(2008)]{Beirao08} Beir{\~a}o, P., et al.\ 2008, \apj, 676, 304 

\bibitem[Beir{\~a}o et al.(2009)]{Beirao09} Beir{\~a}o, P., Appleton, P.~N., Brandl, B.~R., Seibert, M., Jarrett, T., 
\& Houck, J.~R.\ 2009, \apj, 693, 1650

\bibitem[Bernard-Salas et al.(2009)]{Bernard_Salas09} Bernard-Salas, J., et al.\ 2009, \apjs, 184, 230 

\bibitem[Bosch et al.(2009)]{Bosch09} Bosch, G., Terlevich, E., \& Terlevich, R.\ 2009, \aj, 137, 3437

\bibitem[Brandl et al.(2006)]{Brandl06} Brandl, B.~R., et al.\ 2006, \apj, 653, 1129 

\bibitem[Calzetti et al.(2007)]{Calzetti07} Calzetti, D., et al.\ 2007, \apj, 666, 870 

\bibitem[Castor et al.(1975)]{Castor75} Castor, J., McCray, R., \& Weaver, R.\ 1975, \apjl, 200, L107 

\bibitem[Cherchneff(1995)]{Cherchneff95} Cherchneff, I.\ 1995, \apss, 224, 379

\bibitem[Chu \& Mac Low(1990)]{Chu90} Chu, Y.-H., \& Mac Low, M.-M.\ 1990, \apj, 365, 510

\bibitem[Chu \& Kennicutt(1994)]{Chu94} Chu, Y.-H., \& Kennicutt, R.~C., Jr.\ 1994, \apj, 425, 720

\bibitem[Crowther et al.(2010)]{Crowther10} Crowther, P.~A., Schnurr, O., Hirschi, R., Yusof, N., Parker, R.~J., Goodwin, S.~P., \& Kassim, H.~A.\ 2010, \mnras, 1103

\bibitem[Decin et al.(2004)]{Decin04} Decin, L., Morris, P.~W., Appleton, P.~N., Charmandaris, V., Armus, L., \& Houck, J.~R.\ 2004, \apjs, 154, 408 

\bibitem[Deharveng et al.(2005)]{Deharveng05} Deharveng, L., Zavagno, A., \& Caplan, J.\ 2005, \aap, 433, 565

\bibitem[de Koter et al.(1998)]{deKoter98} de Koter, A., Heap, S.~R., \& Hubeny, I.\ 1998, \apj, 509, 879 

\bibitem[Dickel et al.(1994)]{Dickel94} Dickel, J.~R., Milne, D.~K., Kennicutt, R.~C., Chu, Y.-H., \& Schommer, R.~A.\ 1994, \aj, 107, 1067 

\bibitem[Donahue et al.(2011)]{Donahue11} Donahue, M., de Messi{\`e}res, G.~E., O'Connell, R.~W., Voit, G.~M., Hoffer, A., McNamara, B.~R., \& Nulsen, P.~E.~J.\ 2011, \apj, 732, 40

\bibitem[Dopita et al.(2005)]{Dopita05} Dopita, M.~A., et al.\ 
2005, \apj, 619, 755

\bibitem[Dopita et al.(2006a)]{Dopita06a} Dopita, M.~A., et al.\ 2006, \apj, 639, 788 

\bibitem[Dopita et al.(2006b)]{Dopita06b} Dopita, M.~A., et al.\ 
2006, \apj, 647, 244 

\bibitem[Dopita et al.(2006c)]{Dopita06c} Dopita, M.~A., et al.\ 
2006, \apjs, 167, 177 

\bibitem[Draine(2003)]{Draine03} Draine, B.~T.\ 2003, \araa, 41, 241

\bibitem[Fari{\~n}a et al.(2010)]{Farina10} Fari{\~n}a, C., Bosch, G.~L., \& Barb{\'a}, R.~R.\ 2010, IAU Symposium, 266, 391

\bibitem[Feast \& Catchpole(1997)]{Feast97} Feast, M.~W., \& Catchpole, R.~M.\ 1997, \mnras, 286, L1

\bibitem[Fern{\'a}ndez-Ontiveros et al.(2009)]{Fernandez_Ontiveros09} Fern{\'a}ndez-Ontiveros, J.~A., Prieto, M.~A., \& Acosta-Pulido, J.~A.\ 2009, \mnras, 392, L16 

\bibitem[Fischera \& Dopita(2005)]{Fischera05} Fischera, J., \& Dopita, M.\ 2005, \apj, 619, 340

\bibitem[Galliano et al.(2003)]{Galliano03} Galliano, F., Madden, S.~C., Jones, A.~P., Wilson, C.~D., Bernard, J.-P., \& Le Peintre, F.\ 2003, \aap, 407, 159 

\bibitem[Gratton et al.(2004)]{Gratton04} Gratton, R.~G., Bragaglia, A., Clementini, G., Carretta, E., Di Fabrizio, L., Maio, M., \& Taribello, E.\ 2004, \aap, 421, 937

\bibitem[Grebel \& Chu(2000)]{Grebel00} Grebel, E.~K., \& Chu, Y.-H.\ 2000, \aj, 119, 787 

\bibitem[Groves (2004)]{Groves04} Groves, B.\ 2004, "Dust in Photoionized Nebulae", Ph.D. Thesis, Australian National University, Australia

\bibitem[Groves et al.(2008)]{Groves08} Groves, B., Dopita, M.~A., Sutherland, R.~S., Kewley, L.~J., Fischera, J., Leitherer, C., Brandl, B., \& van Breugel, W.\ 2008, \apjs, 176, 438 

\bibitem[Groves et al.(2008)]{Groves08a} Groves, B., Nefs, B., \& Brandl, B.\ 2008, \mnras, 391, L113 

\bibitem[Groves \& Allen(2010)]{Groves10} Groves, B.~A., \& Allen, M.~G.\ 2010, New Astronomy, 15, 614

\bibitem[Haschke et al.(2011)]{Haschke11} Haschke, R., Grebel, E.~K., \& Duffau, S.\ 2011, \aj, 141, 158

\bibitem[Hubeny \& Lanz(1995)]{Hubeny95} Hubeny, I., \& Lanz, T.\ 1995, \apj, 439, 875 

\bibitem[Hunter et al.(1995)]{Hunter95} Hunter, D.~A., Shaya, E.~J., Scowen, P., Hester, J.~J., Groth, E.~J., Lynds, R., \& O'Neil, E.~J., Jr.\ 1995, \apj, 444, 758 

\bibitem[Hunter(1999)]{Hunter99} Hunter, D.~A.\ 1999, New Views of the Magellanic Clouds, 190, 217 

\bibitem[Indebetouw et al.(2006)]{Indebetouw06} Indebetouw, R., et al.\ 2006, Spitzer Proposal ID \#30653, 30653

\bibitem[Indebetouw et al.(2009)]{Indebetouw09} Indebetouw, R., et al.\ 2009, \apj, 694, 84 

\bibitem[Israel et al.(2010)]{Israel10} Israel, F.~P., Wall, W.~F., Raban, D., Reach, W.~T., Bot, C., Oonk, J.~B.~R., Ysard, N., \& Bernard, J.~P.\ 2010, \aap, 519, A67

\bibitem[Johansson et al.(1998)]{Johansson98} Johansson, L.~E.~B., et al.\ 1998, \aap, 331, 857 

\bibitem[Jones et al.(1996)]{Jones96} Jones, A.~P., Tielens, A.~G.~G.~M., Hollenbach, D.~J., \& McKee, C.~F.\ 1996, The Role of Dust in the Formation of Stars, 419 

\bibitem[Kaufman et al.(1999)]{Kaufman99} Kaufman, M.~J., Wolfire, M.~G., Hollenbach, D.~J., \& Luhman, M.~L.\ 1999, \apj, 527, 795 

\bibitem[Kennicutt(1984)]{Kennicutt84} Kennicutt, R.~C., Jr.\ 1984, \apj, 287, 116 

\bibitem[Kennicutt \& Chu(1994)]{Kennicutt94} Kennicutt, R.~C., \& Chu, Y.-H.\ 1994, Violent Star Formation, from 30 Doradus to QSOs, 1

\bibitem[Kennicutt(1998)]{Kennicutt98} Kennicutt, R.~C., Jr.\ 1998, \araa, 36, 189 

\bibitem[Van Kerckhoven et al.(2000)]{Kerckhoven00} Van Kerckhoven, C., et al.\ 2000, \aap, 357, 1013

\bibitem[Kilbinger et al.(2010)]{Kilbinger10} Kilbinger, M., et al.\ 2010, \mnras, 405, 2381

\bibitem[Kim et al.(2007)]{Kim07} Kim, H.-S., et al.\ 2007, \apj, 669, 1003

\bibitem[Kroupa(2002)]{Kroupa02} Kroupa, P.\ 2002, Science, 295, 82 

\bibitem[Lazendic et al.(2003)]{Lazendic03} Lazendic, J.~S., Dickel, J.~R., \& Jones, P.~A.\ 2003, \apj, 596, 287 

\bibitem[Lebouteiller et al.(2008)]{Lebouteiller08} Lebouteiller, V., Bernard-Salas, J., Brandl, B., Whelan, D.~G., Wu, Y., Charmandaris, V., Devost, D., \& Houck, J.~R.\ 2008, \apj, 680, 398

\bibitem[Leitherer et al.(1999)]{Leitherer99} Leitherer, C., et al.\ 1999, \apjs, 123, 3 

\bibitem[Levesque et al.(2010)]{Levesque10} Levesque, E.~M., Kewley, L.~J., \& Larson, K.~L.\ 2010, \aj, 139, 712 

\bibitem[Madden et al.(1999)]{Madden99} Madden, S.~C., Vigroux, L., \& Sauvage, M.\ 1999, The Universe as Seen by ISO, 427, 933

\bibitem[Maercker \& Burton(2005)]{Maercker05} Maercker, M., \& Burton, M.~G.\ 2005, \aap, 438, 663

\bibitem[Malhotra et al.(2001)]{Malhotra01} Malhotra, S., et al.\ 2001, \apj, 561, 766

\bibitem[Massey \& Hunter(1998)]{Massey98} Massey, P., \& Hunter, D.~A.\ 1998, \apj, 493, 180

\bibitem[Meaburn(1984)]{Meaburn84} Meaburn, J.\ 1984, \mnras, 211, 521 

\bibitem[Meixner et al.(2006)]{Meixner06} Meixner, M., et al.\ 2006, \aj, 132, 2268 

\bibitem[Morisset et al.(2004)]{Morisset04} Morisset, C., Schaerer, D., Bouret, J.-C., \& Martins, F.\ 2004, \aap, 415, 577

\bibitem[Oey \& Clarke(1997)]{Oey97} Oey, M.~S., \& Clarke, C.~J.\ 1997, \mnras, 289, 570

\bibitem[Peeters et al.(2004)]{Peeters04} Peeters, E., Mattioda, A.~L., Hudgins, D.~M., \& Allamandola, L.~J.\ 2004, \apjl, 617, L65

\bibitem[Pellegrini et al.(2010)]{Pellegrini10} Pellegrini, E.~W., Baldwin, J.~A., \& Ferland, G.~J.\ 2010, \apjs, 191, 160

\bibitem[Rubio el al.(1998)]{Rubio98}Rubio, M., Barb{\'a}, R.~H., Walborn, N.~R., Probst, R.~G., Garc{\'{\i}}a, J., \& Roth, M.~R.\ 1998, \aj, 116, 1708 

\bibitem[Rubio et al.(2009)]{Rubio09} Rubio, M., Paron, S., \& Dubner, G.\ 2009, \aap, 505, 177

\bibitem[Russell \& Dopita(1992)]{Russell92} Russell, S.~C., \& Dopita, M.~A.\ 1992, \apj, 384, 508

\bibitem[Selman et al.(1999)]{Selman99} Selman, F., Melnick, J., Bosch, G., \& Terlevich, R.\ 1999, \aap, 347, 532 

\bibitem[Siebenmorgen \& Kr{\"u}gel(2007)]{Siebenmorgen07} Siebenmorgen, R., \& Kr{\"u}gel, E.\ 2007, \aap, 461, 445

\bibitem[Silva et al.(1998)]{Silva98} Silva, L., Granato, G.~L., Bressan, A., \& Danese, L.\ 1998, \apj, 509, 103 

\bibitem[Smith et al.(2007)]{Smith07} Smith, J.~D.~T., et al.\ 2007, \pasp, 119, 1133 

\bibitem[Takagi et al.(2003)]{Takagi03} Takagi, T., Arimoto, N., \& Hanami, H.\ 2003, \mnras, 340, 813 

\bibitem[Townsley et al.(2006)]{Townsley06} Townsley, L.~K., Broos, P.~S., Feigelson, E.~D., Brandl, B.~R., Chu, Y.-H., Garmire, G.~P., \& Pavlov, G.~G.\ 2006, \aj, 131, 2140

\bibitem[V{\'a}zquez \& Leitherer(2005)]{Vazquez05} V{\'a}zquez, G.~A., \& Leitherer, C.\ 2005, \apj, 621, 695 

\bibitem[Walborn et al.(1995)]{Walborn95} Walborn, N.~R., MacKenty, J.~W., Saha, A., White, R.~L., \& Parker, J.~W.\ 1995, \apjl, 439, L47 

\bibitem[Walborn \& Blades(1987)]{Walborn87} Walborn, N.~R., \& Blades, J.~C.\ 1987, \apjl, 323, L65

\bibitem[Walborn \& Blades(1997)]{Walborn97} Walborn, N.~R., \& Blades, J.~C.\ 1997, \apjs, 112, 457 

\bibitem[Walborn et al.(2002)]{Walborn02} Walborn, N.~R., Ma{\'{\i}}z-Apell{\'a}niz, J., \& Barb{\'a}, R.~H.\ 2002, \aj, 124, 1601 

\bibitem[Walcher et al.(2011)]{Walcher11} Walcher, J., Groves, B., Budav{\'a}ri, T., \& Dale, D.\ 2011, \apss, 331, 1

\bibitem[Wang(1999)]{Wang99} Wang, Q.~D.\ 1999, \apjl, 510, L139

\bibitem[Werner et al.(1978)]{Werner78} Werner, M.~W., Becklin, E.~E., Gatley, I., Ellis, M.~J., Hyland, A.~R., Robinson, G., \& Thomas, J.~A.\ 1978, \mnras, 184, 365 

\bibitem[Whitney et al.(2008)]{Whitney08} Whitney, B.~A., et al.\ 2008, \aj, 136, 18

\bibitem[Westerlund(1997)]{Westerlund97} Westerlund, B.~E.\ 1997, Cambridge Astrophysics Series, 29

\bibitem[Wolf(2009)]{Wolf09} Wolf, C.\ 2009, \mnras, 397, 520 

\bibitem[Zavagno et al.(2010)]{Zavagno10} Zavagno, A., et al.\ 2010, \aap, 518, L101


\end{thebibliography}
\end{document}